\documentclass[aip,preprint]{revtex4-1}

\pdftrailerid{}

\usepackage[utf8]{inputenc}
\usepackage[T1]{fontenc}
\usepackage{mathtools}
\usepackage{graphicx}
\usepackage{lineno}
\usepackage[a4paper,left=2.5cm,right=2cm,top=3.5cm,bottom=3cm,headsep=16pt,headheight=16pt]{geometry}
\usepackage{xcolor}
\usepackage{hyperref}

\usepackage{mathptmx}
\usepackage{etoolbox}
\usepackage[section]{placeins}
\usepackage{soul}
\usepackage[version=4]{mhchem}
\usepackage{threeparttable}
\usepackage{booktabs}
\usepackage{multirow}

\makeatletter
\newcommand{\warninput}[1]{%
  \filename@parse{#1}%
  \InputIfFileExists{#1}{}{%
    \message{LaTeX Warning: File `%
    \filename@base.\ifx\filename@ext\relax tex\else\filename@ext\fi'%
    not found on input line \the\inputlineno}%
  }%
}
\makeatother

\makeatletter
\def\@email#1#2{%
 \endgroup
 \patchcmd{\titleblock@produce}
  {\frontmatter@RRAPformat}
  {\frontmatter@RRAPformat{\produce@RRAP{*#1\href{mailto:#2}{#2}}}\frontmatter@RRAPformat}
  {}{}
}%
\makeatother

\hypersetup{allcolors=black,colorlinks=true,linktoc=none,pdfcreator={},pdfproducer={}}

\usepackage{amsmath}
\usepackage{amsfonts}
\usepackage{amssymb,amsthm,esint}
\usepackage{fullpage}
\usepackage{bm}
\usepackage{hyperref}
\usepackage{textcomp}
\usepackage{bigints}
\usepackage{dsfont}
\usepackage{colortbl}
\usepackage[normalem]{ulem}
\usepackage{empheq}
\usepackage{dsfont}
\usepackage{comment}
\usepackage{enumitem}
\usepackage[colorinlistoftodos]{todonotes}

\DeclareMathAlphabet\mathbfcal{OMS}{cmsy}{b}{n}

\DeclareMathOperator*{\argmin}{arg\,min}

 \def\R{\mathbb{R}}

 \def\cK{\mathcal{K}}
 
 \def\cN{\mathcal{N}}

\def\bcK{{\mathbfcal{K}}}

\def\bcN{{\mathbfcal{N}}}
\def\bcC{{\mathbfcal{C}}}

\newcommand{\bc}{\mathbf c}

\newcommand{\br}{\mathbf r}

\newcommand{\bR}{\mathbf R}

\usepackage{relsize}

\newcommand{\m}[1]{\texttt{#1}}
\newcommand{\clsn}[1]{\textbf{#1±}}
\newcommand{\clsp}[1]{\textbf{#1+}}
\newcommand{\clsa}[1]{\textbf{a#1}}
\newcommand{\clsg}[1]{\textbf{g#1}}

\newcommand{\clsng}[1]{\textbf{g#1±}}
\newcommand{\clsna}[1]{\textbf{a#1±}}
\newcommand{\clspg}[1]{\textbf{g#1+}}
\newcommand{\clspa}[1]{\textbf{a#1+}}

 \newcommand{\epsOut}{\varepsilon_{\rm out}}
 \newcommand{\epsInn}{\varepsilon_{\rm inn}}
 \newcommand{\opt}{{\rm opt}}

\begin{document}
    \title{Multi-center decomposition of molecular densities: A numerical perspective}

    \author{YingXing Cheng}
    \affiliation{Institute of Applied Analysis and Numerical Simulation, University of Stuttgart, Pfaffenwaldring 57, 70569 Stuttgart, Germany}

    \author{Eric Cancès}
    \affiliation{CERMICS, Ecole des Ponts and Inria Paris, 6 \& 8 Avenue Blaise Pascal, 77455 Marne-la-Vallée, France}

    \author{Virginie Ehrlacher}
    \affiliation{CERMICS, Ecole des Ponts and Inria Paris, 6 \& 8 Avenue Blaise Pascal, 77455 Marne-la-Vallée, France}

    \author{Alston J. Misquitta}
    \affiliation{Queen Mary University of London, Mile End Road, London E1 4NS, United Kingdom}

    \author{Benjamin Stamm} \email{benjamin.stamm@mathematik.uni-stuttgart.de}
    \affiliation{Institute of Applied Analysis and Numerical Simulation, University of Stuttgart, Pfaffenwaldring 57, 70569 Stuttgart, Germany}

    \date{\today}

    \begin{abstract}
    In this study, we analyze various Iterative Stockholder Analysis (ISA) methods for molecular density partitioning, focusing on the numerical performance of the recently proposed Linear approximation of Iterative Stockholder Analysis model (LISA) [J. Chem. Phys. 156, 164107 (2022)].
We first provide a systematic derivation of various iterative solvers to find the unique LISA solution.
In a subsequent systematic numerical study, we evaluate their performance on 48 organic and inorganic, neutral and charged molecules and also compare LISA to two other well-known ISA variants: the Gaussian Iterative Stockholder Analysis (GISA) and Minimum Basis Iterative Stockholder analysis (MBIS).
The study reveals that LISA-family methods can offer a numerically more efficient approach with better accuracy compared to the two comparative methods.
Moreover, the well-known issue with the MBIS method, where atomic charges obtained for negatively charged molecules are anomalously negative, is not observed in LISA-family methods.
Despite the fact that LISA occasionally exhibits elevated entropy as a consequence of the absence of more diffuse basis functions, this issue can be readily mitigated by incorporating additional or integrating supplementary basis functions within the LISA framework.
This research provides the foundation for future studies on the efficiency and chemical accuracy of molecular density partitioning schemes.

    \end{abstract}

    \maketitle


    \section{Introduction}
    \label{sec:introduction}
    In computational chemistry, an interesting question is how to define an atom within a multi-atom molecule.
This plays an important role in many applications.
For example, in the development of traditional force fields,\cite{Case2005,Salomon-Ferrer2013,Salomon-Ferrer2013a,Brooks1983,Brooks2009,Patel2004a,Patel2004,Vanommeslaeghe2015,Zhu2012,Harder2016,Jorgensen1996,Berendsen1995,Pronk2013,VanDerSpoel2005,Phillips2005,Zimmerman1977,Ponder2010,Allinger1989,Allinger2003,Halgren1996,Halgren1999,Hagler1994,Hwang1994,Hwang1998,Maple1994,Maple1994a,Maple1998,Casewit1992,Rappe1992,Lifson1968,Warshel1970,Hagler1979} atoms are usually treated as classical particles with some partial charges, allowing the direct computation of electrostatic interactions.
In polarizable force fields,\cite{Warshel1976,Kaminski2002,Ren2002,Jorgensen2007,Shi2013,Lamoureux2003a,Yu2003,Lemkul2016,Kunz2009,Mortier1985,Mortier1986,Kale2011,Kale2012,Kale2012a,Bai2017,Cools-Ceuppens2022,Verstraelen2013,Verstraelen2014,Cheng2022} these partial charges are also utilized to reproduce molecular polarizabilities.
Partitioning molecules into atomic contributions enables one to define distributed polarizability, and charge-flow contributions to polarizability can then be introduced.\cite{Misquitta2018}
This non-local charge-flow effect could play an important role in non-additive dispersion energy calculations in low-dimensional nanostructures or metallic systems,\cite{Dobson2006,Dobson2012,Misquitta2010,Misquitta2014} where long-range charge fluctuations result in dispersion interactions with non-standard power laws, with a smaller magnitude of the exponent of $R$ where $R$ represents the intermolecular distance.\cite{Dobson2006,Dobson2016}
In addition, recent study shows that the charge-flow effect could also play significant roles in the anisotropy of molecular response properties, e.g., anisotropic dipole polarizability and dispersion coefficients.\cite{Cheng2023a}

The splitting of molecular orbitals/density into atomistic contributions is not an intrinsic property in quantum mechanics.
Therefore, numerous partitioning schemes have been proposed in the literature to calculate atom-in-molecule (AIM) properties.
These methods can be broadly classified into two categories.\cite{Heidar-Zadeh2018}
The first category involves the numerical partitioning of the molecular wavefunction in Hilbert space, such as the orbital-based methods of Mulliken,~\cite{Mulliken1955a,Mulliken1955b,Mulliken1955c,Mulliken1955d} Löwdin,~\cite{Löwdin1950,Löwdin1970,Davidson1967} etc.
The second category divides a molecular descriptor in real space, exemplified by the electron-density-based methods of Hirshfeld,\cite{Hirshfeld1977, Bultinck2007} and Bader,~\cite{Bader1972} etc.
For more details, readers should refer to recent studies.\cite{Ayers2000,C.Lillestolen2008,Lillestolen2009,Bultinck2009,Verstraelen2012b,Verstraelen2013a,Misquitta2014b,Heidar-Zadeh2015,Verstraelen2016,Heidar-Zadeh2018,Benda2022}

In this work, we focus on the real-space methods of the Iterative Stockholder Analysis (ISA) family,\cite{C.Lillestolen2008,Lillestolen2009} which utilize the Kullback–Leibler entropy as the objective functional, with the promolecule density constructed as a sum of spherical, non-negative pro-atom densities, without constraints on their radial dependence.
The partitioning problem is then converted to a constrained optimization problem that provides a mathematically sound definition of an exact partitioning.
For its discretization, recent work by some of our authors proposed a unified framework for ISA family methods from a mathematical perspective and introduced a new scheme, the linear approximation of ISA, denoted as LISA\@.\cite{Benda2022}
The constraint optimization problem defining the LISA-solution is strictly convex and has a unique local minimizer.
This work focuses on the numerical performance of LISA compared to other ISA-based methods.

The remainder of this paper is structured as follows.
In Section~\ref{sec:methods}, we describe the relevant methodology for this study.
We begin by defining relevant spaces and sets and then introduce the global constrained minimization problem defining the LISA-solution.
Either, this problem is solved directly as a global constrained minimization or one can investigate the (non-linear) equations defining critical points of the underlying Lagrangian.
These non-linear equations can be solved iteratively using a fixed point procedure and possibly accelerated using different versions of direct inversion in the iterative subspace method.
Alternatively, these non-linear equations can also be rewritten as a root-finding problem, and therefore,  Newton-type, and quasi-Newton methods can be introduced and utilized.
Furthermore, the problem can also be solved by alternating the minimization between the AIM- and pro-atom densities.
The optimization of the pro-atom densities, for given AIM-densities, has a similar structure as the global, original constrained optimization problem but formulated locally and giving rise to small independent problems which can be solved in parallel.
Nevertheless, all the optimization methods introduced for the global problem above can be employed in this local version.
In Section~\ref{sec:computational-details}, we provide computational details on solver schemes for ISA, including their notation, convergence criteria, and basis functions.
The results and discussions are presented in Section~\ref{sec:res}.
Lastly, a summary is given in Section~\ref{sec:summary}.
Atomic units are used throughout.

    \section{Methods}
    \label{sec:methods}
    We first recall the constrained optimization problem from Ref.~\citenum{Benda2022}, the unique solution of which defines the LISA solution.
We then derive various numerical methods to compute approximations of it.

\subsection{Relevant spaces and sets}
We begin by recalling some notation from a previous article.~\cite{Benda2022}
We introduce the following set for an atomic/molecular density:
\begin{align}
  X := \left\{
    f\in L^1(\R^3)\cap L^\infty(\R^3) \middle| \lim_{|\br|\to \infty} f(\br) = 0, \int_{\R^3} |\br| \, |f(\br)|\, d\br < \infty
  \right\},
\end{align}
and define $X_+ = \{ f\in X \mid f \geq 0 \text{ a.e. } \}$.
Here, $L^1(\R^3)$ indicates that the density $f$ is integrable, and $L^\infty(\R^3)$ means that $f$ is essentially bounded.
The condition $\lim_{|\br|\to \infty} f(\br) = 0$ ensures that the density vanishes at infinite distance, and the condition $\int_{\R^3} |\br| \, |f(\br)|\, d\br < \infty$ ensures that the corresponding dipole moment associated with the density being finite.

We consider a molecule consisting of $M$ atoms and frequently use the variable $a$, where $1 \leq a \leq M$, as the index for the atoms, or more generally, sites.
The convex set $\bcK_{\rho,\bR}$ containing the AIM densities is defined by
\begin{align}
    \bcK_{\rho,\bR}
    =
    \left\{
        \bm \rho = (\rho_a)_{1\le a \le M} \in X_+^M \;\middle|\; \sum_{a=1}^M \rho_a(\br) = \rho(\br)
    \right\},
\end{align}
with $\bR = \{ (\bR_a)_{1 \le a \le M} \in (\R^3)^M \}$, a collection of $M$ sites, and $\rho_a$ being centered at $\bR_a$.
For the LISA discretization, we introduce, for each site $\bR_a$, $m_a$ positive basis functions $g_{a,k} \in X_+$, which are centered at $\bR_a$ and are radially symmetric.
Here, $g_{a,k}:\R^3 \to \R_+$ represents the function in terms of the Cartesian coordinates $\br$, and $\tilde g_{a,k}:\R_+ \to \R_+$ is its radial counterpart, where $\tilde g_{a,k}(r) = g_{a,k}(\br)$ with $r = |\br - \bR_a|$ being the radial distance.
The function $\tilde g_{a,k}(r)$ is monotonically decaying.
Although $\mathbf{R}_a$ could also represent an arbitrary expansion center as pointed out in Ref.~\citenum{Benda2022}, we focus in this work on the case where it denotes the position of the nucleus with index $a$.
Further, we assume that $g_{a,k}(\br)$ is normalized such that
\begin{align}
    \int_{\R^3} g_{a,k}(\br) \, d\br=1,
    \label{eq:norm_condition}
\end{align}
and, we focus on exponential functions of the following format:
\begin{align}
  g_{a,k}(\br) = \frac{n_{a,k} \alpha_{a,k}^{3/n_{a,k}}}{4 \pi \Gamma(3/n_{a,k})}  e^{-\alpha_{a,k} |\br - \bR_a|^{n_{a,k}}}.
  \label{eq:general_exp_basis_func}
\end{align}
The subscripts $a$ and $k$ still denote the indices of atoms and basis functions, respectively.
Specifically, for Gaussian basis functions, we have $n_{a,k} = 2$ for all $a$ and $k$, while for Slater basis functions, $n_{a,k} = 1$ holds true for all $a$ and $k$, corresponding to
\begin{align}
    g_{a,k}(\br) = \left(\frac{\alpha_{a,k}}{\pi}\right)^3 e^{-\alpha_{a,k} |\br-\bR_a|^2},
\end{align}
and
\begin{align}
    g_{a,k}(\br) = \frac{\alpha_{a,k}^3}{8\pi} e^{-\alpha_{a,k} |\br-\bR_a|}
\end{align}
respectively.

For the pro-atom charge distributions, we will now distinguish between
\begin{align}
	\cK^0_{a,\text{LISA}}
	=
	\left\{ \rho_a^0(\br) =\sum_{k=1}^{m_{a}} c_{a, k} \, g_{a,k}(\br), \; c_{a,k} \in \R \right\},
\end{align}
and
\begin{align}
	\cK^0_{a,\text{LISA},+}
	=
	\left\{ \rho_a^0(\br) =\sum_{k=1}^{m_{a}} c_{a, k} \, g_{a,k}(\br), \; c_{a,k} \in \R_+ \right\}.
\end{align}

The difference is that the latter only allows for non-negative coefficients $c_{a,k}$.
In this case, the method becomes similar to the one used in Ref.~\citenum{Tehrani2023} to fit atomic densities.
It is worth noting that the parameters $\alpha_{a,k}$ in Ref.~\citenum{Tehrani2023} are also optimized, whereas in our work, the $\alpha_{a,k}$ values are fixed.

Note that each $\rho_a^0 \in \cK^0_{a,\text{LISA}}$ (or $\cK^0_{a,\text{LISA},+}$) is represented by the vector $\bc_a=(c_{a,1},\ldots,c_{a,m_a}) \in \R^{m_a}$ (or $\R_+^{m_a}$) in terms of
\begin{align}
    \label{eq:rhoa0}
	 \rho_a^0(\br) =\sum_{k=1}^{m_a} c_{a,k} \, g_{a,k}(\br).
\end{align}
We now introduce
\begin{align}
    \bcK^0_{\text{LISA}} &= \cK^0_{1,\text{LISA}} \times \ldots \times \cK^0_{M,\text{LISA}},\\
    \bcK^0_{\text{LISA},+} &= \cK^0_{1,\text{LISA},+} \times \ldots \times \cK^0_{M,\text{LISA},+}.
\end{align}
One main difference in the definition of $\bcK^0_{\text{LISA}}$ (or $\bcK^0_{\text{LISA},+}$) from that in Ref.~\citenum{Benda2022} is the utilization of the same basis functions for identical atom types within a molecule.
However, the definition in this work is more general, as it only incorporates the atom index $a$.
The definition from Ref.~\citenum{Benda2022} can be easily reproduced by assuming that the basis functions are identical for each atom type.

Furthermore, the pro-molecule density, denoted as $\rho^0(\mathbf{r})$, is defined by
\begin{align}
    \label{eq:rho0}
	 \rho^0(\br)
     = \sum_{a=1}^M \rho_a^0(\br)
     = \sum_{a=1}^M \sum_{k=1}^{m_a} c_{a,k} \, g_{a,k}(\br).
\end{align}
All degrees of freedom are then collected in the big vector $\bc = (\bc_1,\ldots,\bc_M) \in \R^P$ (or $\R_+^P$) using encapsulated notation with $P = \sum_{a=1}^M m_a$.
Note that $\rho_a^0(\br)$ depends linearly on $\bc_a$ while $\rho^0(\br)$ depends linearly on $\bc$.

For two sets of AIM and pro-atom densities $\bm\rho\in\bcK_{\rho,\bR}$ and $\bm\rho^0\in \bcK^0_{\text{LISA}}$ (or $\bcK^0_{\text{LISA},+}$), we now introduce the relative entropy given by the Kullback-Leibler (KL) divergence\cite{Benda2022}
\begin{align}
    \label{eq:entropy}
    S(\bm\rho, \bm\rho^0)
    = \sum_{a=1}^M \int_{\R^3} \rho_a(\br) \ln\left(\frac{\rho_a(\br)}{\rho_a^{0}(\br)}\right) \, d\br,
\end{align}
with the conventions
\begin{align}
    0\cdot \ln\left(\frac{0}{0}\right)&=0                   &
    p\cdot\ln\left(\frac{p}{0}\right) &=\infty \quad \forall p>0 &
    0\cdot\ln\left(\frac{0}{p}\right) &=0.
\end{align}

In this work, we focus exclusively on the KL divergence; however, other objective functions are also available,\cite{Heidar-Zadeh2018} such as the quadratic error integral function used in the Gaussian iterative stockholder analysis (GISA) model,\cite{Verstraelen2012b} and the Atomic Shell Approximation (ASA) model.\cite{Constans1995}

\subsection{Definition of the LISA solution}
We now slightly deviate from the presentation of LISA introduced in Ref.~\citenum{Benda2022} since we will present two different, but highly connected, variants.
Indeed, we consider first the version introduced in Ref.~\citenum{Benda2022}
\begin{align}
    \label{eq:general-pos}
    \left( \bm\rho^{\opt}_+, \bm\rho_+^{0, \opt}\right) \in \mathop{\rm argmin}_{\left( \bm\rho, \bm\rho^{0}\right) \in\bcC_{\rho,\bR,+} } S(\bm\rho, \bm\rho^0)
\end{align}
where $\bcC_{\rho,\bR,+}$ is given by
\begin{align}
	\bcC_{\rho,\bR,+}= \big\{  (\bm\rho,\bm\rho^0) \in \bcK_{\rho,\bR} \times  \bcK^0_{\text{LISA},+} \; \big| \; \bcN(\bm\rho)=\bcN(\bm\rho^0) \big\},
\end{align}
and $\bcN(\bm\rho)$ denotes the vector of $M$ components given by
\begin{align}
    [\bcN(\bm\rho)]_a = \cN(\rho_a) = \int_{\R^3} \rho_a(\br) \, d\br  \qquad \forall \, 1 \le a \le M.
\end{align}

Second, we introduce the variant without non-negativity condition on the coefficients $\bc \in \R^P$, i.e.
\begin{align}
    \label{eq:general}
    \left( \bm\rho^{\opt}, \bm\rho^{0, \opt}\right) \in \mathop{\rm argmin}_{\left( \bm\rho, \bm\rho^{0}\right) \in\bcC_{\rho,\bR} } S(\bm\rho, \bm\rho^0)
\end{align}
where $\bcC_{\rho,\bR}$ is given by
\begin{align}
    \bcC_{\rho,\bR}= \big\{  (\bm\rho,\bm\rho^0) \in \bcK_{\rho,\bR} \times  \bcK^0_{\text{LISA}} \; \big| \; \bcN(\bm\rho)=\bcN(\bm\rho^0) \big\}.
\end{align}
Note that we have three kinds of constraints in both variants:
\begin{itemize}
    \item[] (C1):
    Decomposition of the charge:
    \begin{align}
        \sum_{a=1}^M \rho_a(\br) = \rho(\br)
        \label{eq:constraint_c1}
    \end{align}
    \item[] (C2):
    Consistency between AIM- and pro-atom charge: $\bcN(\bm\rho) = \bcN(\bm\rho^0)$, i.e.,
    \begin{align}
        \int_{\R^3} \rho_a(\br)\, d\br  = \int_{\R^3} \rho^0_a(\br)\, d\br \qquad \forall a=1,\ldots,M.
        \label{eq:constraint_c2}
    \end{align}
    The extra degrees of freedom enable one to ensure that the population of each pro-atom and its corresponding AIM are equal, thus eliminating any ambiguity in the statistical interpretation of Eq.~\eqref{eq:entropy}.\cite{Parr2005}
    \item [] (C3):
    Positivity of the pro-atom density:
    \begin{align}
        \rho_a^0(\br) \ge 0 \qquad \forall \, \br \in \R^3,
        \quad \text{or} \quad
        \sum_k c_{a,k} \tilde{g}_{a,k}(r) \ge 0 \qquad \forall \, r \in \R_+.
        \label{eq:constraint_c3}
    \end{align}
    It should be noted that if $S(\bm{\rho}, \bm{\rho}^0) < +\infty$ holds, then condition (C3) is always satisfied, because a locally negative pro-atom density leads to infinite entropy.
    However, this condition can be applied to create a robust implementation, for example, by improving the Newton method discussed below through the addition of a line search.
\end{itemize}
Additionally, we note that the following property has been discussed in the chemistry community: the spherical average of $\rho_a^0(\br)$, denoted as $\rho_a^0(r)$ with $r=|\br-\bR_a|$ , should decay monotonically,\cite{Weinstein1975,Simas1988,Ayers2003,Verstraelen2012b,Heidar-Zadeh2018}
\begin{align}
    \frac{\partial{\rho_a^0(r)}}{\partial{r}} \le 0, \quad \text{or} \quad
    \sum_k c_{a,k} \frac{\partial \tilde{g}_{a,k}(r)}{\partial r} \le 0.
    \label{eq:constraint_cp}
\end{align}
This chemical constraint is important in practice.
Note that the condition expressed in Eq.~\eqref{eq:constraint_cp} is always satisfied by the solution to Eq.~\eqref{eq:general-pos}, given that $\tilde{g}_{a,k}$ takes the form described in Eq.~\eqref{eq:general_exp_basis_func}.
This is because each $g_{a,k}$ is decreasing and each $c_{a,k}$ is non-negative.
However, numerical implementations of the ISA that use grids encounter difficulties at the chemical level when a central atom is encased in a spherical shell of other atoms.
In such cases, the numerical implementations of ISA lead to significantly increase at the locations of subsequent atom shells.\cite{Verstraelen2012b,Heidar-Zadeh2018}
This can result to an abnormally high population for the central atom, resulting in an atomic density that is nonmonotonic and violates the ``sensibility'' requirement.\cite{Heidar-Zadeh2018}
ISA also lacks conformational stability, as even a minor alteration of the molecular symmetry (where atoms surrounding the central atom are positioned on an elliptical rather than a spherical surface) can affect the density.\cite{Heidar-Zadeh2018}

The solution $ \bm\rho^{0, \opt}_+$ of Eq.~\eqref{eq:general-pos} satisfies all Eqs.~\eqref{eq:constraint_c1}-\eqref{eq:constraint_cp}  since, in $\cK^0_{a,\text{LISA},+}$, there holds $c_{a,k}\ge 0$, $g_{a,k}(\br)\ge 0$ and $\frac{\partial{\tilde g_{a,k}(r)}}{\partial{r}} \le 0$.
Note that there can be solutions with negative coefficients that still satisfy Eqs.~\eqref{eq:constraint_c1}-\eqref{eq:constraint_c3}.
In this sense, restricting the pro-atom densities to $\bcC_{\rho,\bR}$ is sufficient for Eq.~\eqref{eq:constraint_cp} to hold but not necessary.
Further, since $\bcC_{\rho,\bR,+}\subset \bcC_{\rho,\bR}$ the entropy might be higher:
\begin{align}
    S(\bm\rho^{\opt}, \bm\rho^{0,\opt})
    \le
    S(\bm\rho^{\opt}_+, \bm\rho^{0,\opt}_+).
\end{align}

As pointed out in Ref.~\citenum{Verstraelen2016} and Remark 3 of Ref.~\citenum{Benda2022}, it can be proven that the LISA-problems Eqs.~\eqref{eq:general-pos} and~\eqref{eq:general} are equivalent to the alternative constrained optimization problem with Eq.~\eqref{eq:rho0}:
\begin{align}
    \min_{\bm{c} \in \R_+^P}
    \int_{\R^3} \rho(\br) \ln\left( \frac{\rho(\br)}{\rho^0(\br)} \right) d\br,
    \qquad \text{s.t.} \qquad \text{Eq.}~\eqref{eq:common-constraint}
    \label{eq:alt-const-opt-pos}
\end{align}
and
\begin{align}
    \min_{\bm{c} \in \R^P}
    \int_{\R^3} \rho(\br) \ln\left( \frac{\rho(\br)}{\rho^0(\br)} \right) d\br,
    \qquad \text{s.t.} \qquad \text{Eq.}~\eqref{eq:common-constraint}
    \label{eq:alt-const-opt}
\end{align}
with
\begin{equation}
    \int_{\R^3} (\rho^0(\br) - \rho(\br)) \, d\br = 0.
    \label{eq:common-constraint}
\end{equation}
%

The atomic densities $\rho_a$ are then obtained by,
\begin{align}
    \rho_a(\br) &= \rho(\br)  w_a(\br),
    \label{eq:rho_a_by_def}
    \\
    w_a(\br) &= \frac{\rho_a^0(\br)}{\rho^0(\br)},
    \label{eq:aim_weight_func}
\end{align}
in which the functions $w_a(\bm{r})$ ($0 \leq w_a(\bm{r}) \leq 1$) are so-called AIM weights functions, determining which proportion of the molecular density is assigned to atom $a$. \cite{Stone1981}

In the following, we present different approaches to compute the LISA approximation.
These approaches are derived from one of constrained optimization problems presented by Eqs.~\eqref{eq:general-pos},~\eqref{eq:general},~\eqref{eq:alt-const-opt-pos}-\eqref{eq:alt-const-opt}.
It should be noted that in the practical context of implementing the methods, any integral will be replaced by appropriate quadrature that will be specified in Section~\ref{sec:computational-details}.
For sake of a simple presentation, we present here all methods with exact quadrature.

\subsection{Global approach} 
\label{ssec:GlobalLagrangain}
We first investigate the optimality conditions of the convex constrained optimization problem Eq.~\eqref{eq:alt-const-opt}.
Since this is a convex optimization problem in the discrete variable $\bc$ with an affine constraint, let us thus introduce the Lagrangian
\begin{align}
     L_{{\rm glob}}(\bc,\mu) = \int_{\R^3} \rho(\br) \ln\left( \frac{\rho(\br)}{\rho^0(\br)} \right) d\br + \mu  \int_{\R^3} \left(\rho^0(\br) - \rho(\br)\right) d\br,
    \label{eq:global_lagrangian_mu}
\end{align}
where the dependency of $\bc$ enters through $\rho^0$ defined in Eq.~\eqref{eq:rho0}.
Differentiating with respect to $c_{a,k}$ yields
\begin{align}
  0 = \frac{\partial  L_{{\rm glob}}(\bc,\mu)}{\partial c_{a,k}}
  = - \int_{\R^3} \frac{\rho(\br)}{\rho^0(\br)} g_{a,k}(\br)\, d\br
  + \mu \int_{\R^3} g_{a,k}(\br) \, d\br.
\end{align}
Multiplying by $c_{a,k}$ and summing over all $a,k$ yields
\begin{align}
  \int_{\R^3} \rho(\br) \, d\br = \mu \int_{\R^3} \rho^0(\br) \, d\br.
\end{align}
Since
\begin{align}
  \frac{\partial  L_{{\rm glob}}(\bc, \mu)}{\partial \mu} = 0
  \qquad
  \Leftrightarrow
  \qquad
  \int_{\R^3} \rho^0(\br)\,
  d\br = \int_{\R^3} \rho(\br) \, d\br,
\end{align}
we deduce that $\mu=1$ and, since $\int_{\R^3} g_{a,k}(\br) \, d\br=1$, that
the unique solution of Eq.~\eqref{eq:alt-const-opt} satisfies
\begin{align}
    \label{eq:Global-EL}
    \int_{\R^3} \frac{\rho(\br)}{\rho^0(\br)} g_{a,k} (\br) \,d\br = 1.
\end{align}

Multiplying Eq.~\eqref{eq:Global-EL} by $c_{a,k}$ and summing over $k$, in combination with Eqs.~\eqref{eq:norm_condition} and \eqref{eq:rho_a_by_def}, yields
\begin{align}
  \int_{\R^3} \rho_a^0(\br) \, d\br
  = \sum_{k} c_{a,k} \int_{\R^3} g_{a,k}(\br) \, d\br
  = \int_{\R^3} \rho(\br) \frac{\rho_a^0(\br)}{\rho^0(\br)} d\br
  = \int_{\R^3} \rho_a(\br) \, d\br,
\end{align}
satisfying thus (C2) as well.

In the following, we now present different solvers either based on the optimality condition Eq.~\eqref{eq:Global-EL} or on the direct convex constrained optimality problem Eqs.~\eqref{eq:alt-const-opt-pos}-\eqref{eq:alt-const-opt}.

\subsubsection{Fixed-point iterations and accelerations thereof}
Multipliying Eq.~\eqref{eq:Global-EL} by $c_{a,k}$ gives rise to a natural fixed-point iteration scheme
\begin{align}
    \label{eq:fixed-point-global}
    c^{(m+1)}_{a,k}
     =
    \int_{\R^3} \frac{\rho(\br)}{\rho^{0,(m)}(\br)} c^{(m)}_{a,k}g_{a,k}(\br) \, d\br.
\end{align}
with
\begin{align}
    \rho^{0,(m)}(\br) = \sum_{a,k} c^{(m)}_{a,k}g_{a,k}(\br).
    \label{eq:rho0_m}
\end{align}
Note that this iterative scheme conserves the sign of $c_{a,k}$, i.e. if a non-negative initial condition $c^{(0)}_{a,k}$ is chosen, there holds $c^{(m)}_{a,k} \ge 0$ for all iterations $m$.
In this sense, this scheme, if initialized with non-negative $c_{a,k}$, does not allow solving the minimization problem Eq.~\eqref{eq:general} if the solution contains a negative coefficient $c_{a,k}$.
We refer to Eq.~\eqref{eq:fixed-point-global} as \m{gLISA-SC} where ``SC'' stands for self-consistency.

This fixed-point iterative scheme can, in theory, be accelerated using different versions of direct inversion in the iterative subspace (DIIS).
It should be noted that, although DIIS has been demonstrated to be equivalent to the quasi-Newton method,\cite{Eyert1996,Fang2009,Walker2011,Chupin2021} which will be discussed in Section~\ref{sss:quasi_newton_methods}, we treat it here as an acceleration method for the fixed-point problem.
However, for the non-negative solution in Eq.~\eqref{eq:alt-const-opt-pos}, it is not straightforward to preserve positivity due to the mixing process.
In this work, we utilize the DIIS methods as proposed in Ref.~\citenum{Chupin2021}, albeit with two modifications.
First, the residual function is redefined as the deviation between the optimization solutions of the current and previous iterations, as opposed to using the commutator from Ref.~\citenum{Chupin2021}, which was specifically designed for computational chemistry applications.
Consequently, the methods employed in this work include restarted DIIS (R-DIIS), fixed-depth DIIS (FD-DIIS), and adapted-depth DIIS (AD-DIIS), corresponding to R-CDIIS, FD-CDIIS, and AD-CDIIS in Ref.\citenum{Chupin2021}, respectively.
Second, we apply an upper bound to the DIIS subspace size, a step that was not necessary in the original work where the number of unknown parameters, i.e., the elements of the Fock matrix, is typically larger than the DIIS size.
The  gLISA methods with R-DIIS, FD-DIIS, and AD-DIIS are denoted as \m{ gLISA-R-DIIS}, \m{ gLISA-FD-DIIS}, and \m{ gLISA-AD-DIIS}, respectively.
All three methods are designed, in principle, to solve the solution as described in Eq.~\eqref{eq:alt-const-opt}.

\subsubsection{Newton Method}
\label{sss:newton_methods}

The Newton method can be applied to both unconstrained optimization and root-finding problems within the context of our study.
Indeed, Eq.~\eqref{eq:Global-EL} is a root-finding problem where we seek the roots of the vector function $\bm{h}(\bm{c}) = \bm{0}$, with each component defined as:
\begin{align}
    h_{a,k}(\bm{c})
    &=
    1 - \int_{\mathbb{R}^3} \frac{\rho(\mathbf{r})}{\rho^0(\mathbf{r})} g_{a,k}(\mathbf{r}) \, d\mathbf{r}.
    \label{eq:newton_h}
\end{align}

On the other hand, since the critical point of the Lagrangian in Eq.~\eqref{eq:global_lagrangian_mu} is of the form ($\bc_\text{opt}, \mu=1$), $\bc_\text{opt}$ is also the unique minimum of $F_\text{glob}$ defined by
\begin{align}
    F_\text{glob}(\mathbf{c})
    &=
    \int_{\mathbb{R}^3} \rho(\mathbf{r}) \ln\left(\frac{\rho(\mathbf{r})}{\rho^{0}(\mathbf{r})}\right) \, d\mathbf{r}
    + \int_{\mathbb{R}^3} \left( \rho^{0}(\mathbf{r}) - \rho(\mathbf{r})\right) \, d\mathbf{r}.
    \label{eq:global_lagrangian}
\end{align}

This formulation leads to an unconstrained optimization problem that can be solved using Newton method, which involves computing the gradient and Hessian of $ F_\text{glob}(\mathbf{c})$ as follows:
\begin{align}
    \frac{\partial  F_\text{glob}(\mathbf{c})}{\partial c_{a,k}} &=
    1 - \int_{\mathbb{R}^3} \frac{\rho(\mathbf{r})}{\rho^0(\mathbf{r})} g_{a,k}(\mathbf{r}) \, d\mathbf{r}
    = h_{a,k}(\bm{c}),
    \label{eq:newton_gradient} \\
    \frac{\partial^2  F_\text{glob}(\mathbf{c})}{\partial c_{a,k} \partial c_{a',k'} }
    &= \int_{\mathbb{R}^3} \frac{\rho(\mathbf{r})}{[\rho^0(\mathbf{r})]^2} g_{a,k}(\mathbf{r}) g_{a',k'}(\mathbf{r}) \, d\mathbf{r}
    = Dh_{(a,k),(a',k')}(\bm{c}).
    \label{eq:newton_hessian}
\end{align}

The Newton step in this context is formulated as $\bm{c}^{(m+1)} = \bm{c}^{(m)} + \delta^{(m)}$, where $\delta^{(m)}$ is obtained by solving the linear system involving the Jacobian matrix $D\bm{h}(\bm{c}^{(m)})$ of $\bm{h}$ defined by Eq.~\eqref{eq:newton_hessian}.

However, the Newton method might not always respect the non-negativity constraints for some variables during the iterations.
To address this, we implement a modified Newton method (referred to as \m{ gLISA-M-NEWTON}), which includes step-size control to ensure that $\rho_a^0(\mathbf{r}) \ge 0$ at all quadrature points. This approach guarantees to comply with the chemical constraints of the problem.

While we employ Newton method only for Eq.~\eqref{eq:alt-const-opt}, the step-size control could also be adapted for Eq.~\eqref{eq:alt-const-opt-pos} and the constraint $c_{a,k}\ge 0$.

\subsubsection{Quasi-Newton Method}
\label{sss:quasi_newton_methods}
The primary challenge in applying the Newton method to unconstrained optimization problems is computing the Hessian matrix, which is particularly computationally expensive for large-scale problems.
To circumvent this, the quasi-Newton method provides an efficient alternative by approximating the (inverse) Hessian matrix, thus reducing computational overhead.

In this study, we use the Broyden-Fletcher-Goldfarb-Shanno (BFGS) algorithm, a widely recognized quasi-Newton method.\cite{Nocedal2006}
%
This approach efficiently approximates the (inverse) Hessian matrix without direct computation, rendering the quasi-Newton method a preferred choice for handling large-scale optimization tasks.

Initially, we set the approximated (inverse) Hessian matrix to the identity matrix to commence the optimization process.
The step size $\alpha_m$ is dynamically determined through a line search technique, akin to the one used in the modified Newton method, ensuring optimal progression along the descent path.
Consequently, this method is denoted as \m{gLISA-QUASI-NEWTON} in our framework.

While the primary application of the quasi-Newton method in this study is directed at solving Eq.~\eqref{eq:alt-const-opt}, it is important to note that the step-size control mechanism is adaptable and can be extended to handle the positivity constraint $c_{a,k} \ge 0$ in Eq.~\eqref{eq:alt-const-opt-pos}, ensuring that the solution remains chemically viable throughout the optimization process.

\subsubsection{Convex minimization method}
\label{sssec:DirecMinimization}
Finally, we consider a direct minimization approach.
Eqs. \eqref{eq:alt-const-opt-pos} can be treated as a convex optimization problem where the objective function is
\begin{align}
    s_\text{KL}(\bc)=\int_{\mathbb{R}^3} \rho(\mathbf{r}) \ln\left( \frac{\rho(\mathbf{r})}{\rho^0(\mathbf{r})} \right) d\mathbf{r},
    \label{eq: gLISA_convex_obj_func}
\end{align}
with equality constraints defined in Eq.~\eqref{eq:common-constraint} and inequality constraints, i.e., $\mathbf{c} \ge 0$.
Therefore, this constitutes a strictly convex optimization problem.
The gradient and Hessian of $s_\text{KL}(\bc)$ are given by
\begin{align}
    \frac{\partial s_\text{KL}(\bc)}{c_{a,k}} &=
    - \int_{\R^3} \frac{\rho(\br)}{\rho^0(\br)} g_{a,k}(\br) \, d\br
    \label{eq: gLISA_convex_gradient}
    \\
    \frac{\partial^2 s_\text{KL}(\bc)}{\partial c_{a,k} \partial c_{a',k'} }
    &= \int_{\R^3} \frac{\rho(\br)}{[\rho^0(\br)]^2} g_{a,k}(\br) g_{a',k'}(\br) \, d\br.
    \label{eq: gLISA_convex_hessian}
\end{align}
We refer to this method as \m{ gLISA-CVXOPT}.

\bigskip

Note that all methods introduced so far have the prefix ``gLISA-'' which refers to iterative procedures to solve the LISA-problem directly based on a global (\m{g}) ansatz either by direct minimization of Eq.~\eqref{eq:alt-const-opt-pos},~\eqref{eq:alt-const-opt} or based on the global optimality condition Eq.~\eqref{eq:Global-EL}.

\subsection{Alternating minimization methods}
\label{ssec:AlternatingMinimization}
As proposed in Ref.~\citenum{Benda2022}, to solve the solution to Eq.~\eqref{eq:general-pos} or~\eqref{eq:general} we can also consider the following alternating minimization scheme:
\begin{itemize}
    \item[] \textbf{Initialization:}
    Let $\bcK^0 = \bcK^0_{\mathrm{\text{LISA},+}}$ or $\bcK^0 = \bcK^0_{\mathrm{\text{LISA}}}$ depending on whether one aims to solve Eq.~\eqref{eq:general-pos} or Eq.~\eqref{eq:general}.
    Choose $\bm{\rho}^{0,(0)} \in \bcK^0$ such that $S(\bm{\rho}|\bm{\rho}^{0,(0)}) < +\infty$.

    \item[] \textbf{Iteration $m \geq 1$:}
    \begin{itemize}
        \item[] \textbf{Step 1:}
        Set
        \begin{align}
            \label{eq:step1}
            \bm{\rho}^{(m)} = \argmin_{\bm{\rho} \in \bcK_{\rho,\bR}} ~ S(\bm{\rho}|\bm{\rho}^{0,(m-1)}),
        \end{align}

        \item[] \textbf{Step 2:}
        Find
        \begin{align}
            \label{eq:step2}
            \bm{\rho}^{0,(m)} \in \argmin_{\substack{\bm{\rho}^0 \in \bcK^0, \\ \bcN(\bm{\rho}^0) = \bcN(\bm{\rho}^{(m)})}} S(\bm{\rho}^{(m)}|\bm{\rho}^0).
        \end{align}
    \end{itemize}
\end{itemize}

The solution to step 1, i.e., Eq.~\eqref{eq:step1}, is given by\cite{Benda2022}
\begin{align}
    \label{eq:step1-explict}
    \rho_a^{(m)} = \rho(\br) w_a^{(m-1)}(\br)
\end{align}
where $w_a^{(m-1)}$ is the $(m-1)$-th iteration AIM weights functions of atom $a$.
It remains to clarify Step 2 given by Eq.~\eqref{eq:step2}.
Since the entropy $S$ is a sum over local (i.e. site-wise) contributions, see Eq.~\eqref{eq:entropy}, and the constraints are also local, the constrained optimization problems Eq.~\eqref{eq:step2} can be solved independently for each $a$ and writes
\begin{align}
    \label{eq:local-step2-pos}
    \min_{
        \rho_a^0 \in \cK^0_{\rm \text{LISA},+}
    }
    \int_{\R^3} \rho_a^{(m)}(\br)
    \ln
    \left(
        \frac{\rho_a^{(m)}(\br)}{\rho_a^{0}(\br)}
    \right)\,
    d\br,
\end{align}
and
\begin{align}
    \min_{
        \rho_a^0 \in \cK^0_\text{LISA}
    }
    \int_{\R^3} \rho_a^{(m)}(\br)
    \ln
    \left(
        \frac{\rho_a^{(m)}(\br)}{\rho_a^{0}(\br)}
    \right)\,
    d\br,
    \label{eq:local-step2}
\end{align}
corresponding to $\bcK^0 = \bcK^0_{\mathrm{\text{LISA},+}}$ and $\bcK^0 = \bcK^0_{\mathrm{\text{LISA}}}$, respectively,
subject to the constraint (C2), i.e., Eq.~\eqref{eq:constraint_c2}.

These independent problems can be solved by different means, either by direct minimization taking the constraints into
account or by solving the resulting (local) non-linear equations defining the critical point(s) of a (local) Lagrangian.

It should be noted that for a single-atom molecule, Eqs.~\eqref{eq:local-step2-pos}-\eqref{eq:local-step2} can be treated as special cases of Eqs.~\eqref{eq:alt-const-opt-pos}-\eqref{eq:alt-const-opt} with $m=0$, respectively.
In addition, $\rho_a^{(m)}$ is provided in each iteration.
In this sense, the notion ``atom-in-molecule`` has a second meaning for this class of solvers.
Therefore, all methods proposed in the global approach can be used in the local approach as well.
Next, we provide more mathematical details for the numerical implementations.

One way to compute the minimum is by solving the first-order optimality condition of this constrained optimization problem.
To do this, we introduce the (local) Lagrangian $L_{{\rm loc},a}$ associated with Eq.~\eqref{eq:local-step2} as follows:
\begin{align}
	 L_{{\rm loc},a}(\bc_a,\mu_a)
	=
	\int_{\R^3} \rho_a^{(m)}(\br) \ln\left(\frac{\rho_a^{(m)}(\br)}{\rho_a^{0}(\br)}\right)\, d\br
	+
	\mu_a \int_{\R^3} \left[ \rho_a^{0}(\br) - \rho_a^{(m)}(\br)\right] \, d\br,
\label{eq:local_lagrangian_mu}
\end{align}
where we remind that $\rho_a^{0}$ depends (linearly) on $\bc_a$ through Eq.~\eqref{eq:rhoa0}.
The computation of the first-order optimality condition is then as in Eq.~\eqref{eq:Global-EL} and writes
\begin{align}
\label{eq:1st-opt-1}
0
& = - \int_{\R^3} \frac{\rho_a^{(m)}(\br)}{\rho_a^{0}(\br)}  g_{a,k}(\br) \, d\br + \mu_a.
\end{align}

Analogously to the developments of Section~\ref{ssec:GlobalLagrangain}, one can show that  $\mu_a=1$.
The first-order optimality condition of the constrained optimization problem Eq.~\eqref{eq:local-step2} then writes
\begin{align}
    \label{eq:local-opt-cond}
    1 = \int_{\R^3} \frac{\rho_a^{(m)}(\br)}{\rho^0_a({\br})} g_{a,k}(\br) \, d\br .
\end{align}
For each $a$, this is a set of $m_a$ coupled non-linear equations in $c_{a,k}$, which can be solved using different techniques that are explained in the following.

\subsubsection{Local fixed-point iterations}
Multiplying Eq.~\eqref{eq:local-opt-cond} by $c_{a,k}$ provides a basis to define a (local) fixed-point iteration
\begin{align}
	c^{(m,\ell+1)}_{a,k}
	=
    \int_{\R^3} \frac{\rho_a^{(m)}(\br)}{\rho_a^{0,(m,\ell)}(\br)}  c^{(m,\ell)}_{a,k} g_{a,k}(\br) \, d\br.
    \label{eq:lisa_sc_eq}
\end{align}
with
\begin{align}
    \rho_a^{0,(m,\ell)}(\br ) = \sum_{k=1}^{m_{z_a}} c^{(m,\ell)}_{a,k} \, g_{a,k}(\br).
    \label{eq:lisa_sc_rho0_a}
\end{align}
We refer to this method as \m{aLISA-SC}.
We observe that, with a non-negative initialization $c^{(m,0)}_{a,k} = c^{(m-1,\ell_{\max})}_{a,k} \geq 0$, this scheme conserves non-negativity, similar to its global variant \m{gLISA-SC}, and is thus intended to solve Eq.~\eqref{eq:local-step2}.

The remark concerning the non-negativity of the coefficients in the global iteration schemes in Eq.~\eqref{eq:fixed-point-global} also applies to the local iteration schemes in Eq.~\eqref{eq:lisa_sc_eq}.
Again, this fixed-point iterative scheme can be accelerated using DIIS, yielding the methods \m{aLISA-R-DIIS}, \m{aLISA-FD-DIIS} and \m{aLISA-AD-DIIS}.
We observe again that no sign constraint is imposed on these methods due to the mixing; thus, this scheme aims to solve Eq.~\eqref{eq:local-step2}.

\subsubsection{Local Newton method}
We can use exactly the same arguments as explained in Section~\ref{sss:newton_methods}, but on a local level to convert the (local) constrained optimization problem into an unconstrained one.
Thus, we consider the following objective function:
\begin{align}
	 F_{{\rm loc},a}(\bc_a)
	=
	\int_{\R^3} \rho_a^{(m)}(\br) \ln\left(\frac{\rho_a^{(m)}(\br)}{\rho_a^{0}(\br)}\right)\, d\br
	+
	\int_{\R^3} \left[ \rho_a^{0}(\br) - \rho_a^{(m)}(\br)\right] \, d\br,
    \label{eq:merged_lagrangian}
\end{align}
the unique local minimizer of which coincides with the solution of Eq.~\eqref{eq:local-step2}.
The gradient and Hessian of $F_{{\rm loc},a}(\bc_a)$ are crucial for applying Newton method:
\begin{align}
    \frac{\partial  F_{{\rm loc},a}(\bc_a)}{\partial c_{a,k}} &=
    1 - \int_{\R^3} \frac{\rho_a^{(m)}(\br)}{\rho_a^0(\br)} g_{a,k}(\br) \, d\br
    =:h_{(a,k)}(\bm{c}_a), \\
    \frac{\partial^2  F_{{\rm loc},a}(\bc_a)}{\partial c_{a,k} \partial c_{a',k'} }
    &= \int_{\R^3} \frac{\rho_a^{(m)}(\br)}{[\rho_a^0(\br)]^2} g_{a,k}(\br) g_{a',k'}(\br) \, d\br
    =:  Dh_{(a,k),(a,k')}(\bm{c}_a).
\end{align}

The update step $\bm{c}_a^{(m,\ell+1)} = \bm{c}_a^{(m,\ell)} + \delta_a^{(m,\ell)}$ involves the Jacobian matrix $D\bm{h}_a(\bm{c}_a)$:
\begin{align}
    D\bm{h}_a(\bm{c}_a^{(m,\ell)}) \, \delta_a^{(m,\ell)} = - \bm{h}_a(\bm{c}_a^{(m,\ell)}).
    \label{eq:lisa_newton_for_eq_jacobian}
\end{align}

Again, we apply a step-size control to maintain the constraint $\rho_a^0(r)\ge 0$ at each quadrature point and refer to this method as \m{aLISA-M-NEWTON}.

\subsubsection{Local quasi-Newton method}
In the same spirit as for \m{gLISA-QUASI-NEWTON}, one can also use a quasi-Newton method for solving Eq.~\eqref{eq:local-step2}.
The BFGS method employed in \m{gLISA-QUASI-NEWTON} can also be used to obtain the approximated (inverse) Hessian matrix, and we refer to this method as \m{aLISA-QUASI-NEWTON} intending to solve Eq.~\eqref{eq:local-step2}.

\subsubsection{Local convex optimization}
Finally, analogous to the global solvers and as explained in Ref.~\citenum{Benda2022}, we also present the strategy to solve the directly constrained optimization problem Eq.~\eqref{eq:local-step2-pos} under the non-negativity constraint in $\cK^0=\cK^0_{\rm LISA,+}$.
The objective function is given by
\begin{align}
    s_\text{KL}(\bc_a) = \int_{\R^3} \rho_a^{(m)}(\br) \ln\left(\frac{\rho_a^{(m)}(\br)}{\rho_a^{0}(\br)}\right)\, d\br,
    \label{eq:lisa_convex_obj_func}
\end{align}
and gradient and Hessian of $s_\text{KL}(\bc_a)$ are given by
\begin{align}
    \frac{\partial s_\text{KL}(\bc_a)}{c_{a,k}} &=
    - \int_{\R^3} \frac{\rho(\br)}{\rho^0(\br)} g_{a,k}(\br) \, d\br
    \label{eq:lisa_convex_gradient}
    \\
    \frac{\partial^2 s_\text{KL}(\bc_a)}{\partial c_{a,k} \partial c_{a',k'} }
    &= \int_{\R^3} \frac{\rho(\br)}{[\rho^0(\br)]^2} g_{a,k}(\br) g_{a',k'}(\br) \, d\br.
    \label{eq:lisa_convex_hessian}
\end{align}
We refer to this method as \m{aLISA-CVXOPT}.

\bigskip

We finish this subsection with the remark that all variants of methods originating from the alternating minimization scheme have a similar numerical performance since computing the AIM-densities by Eq.~\eqref{eq:step1-explict} is the far-most computationally expensive part as it is the only non-local computation and the number of outer iterations (indexed by $m$) is independent of the local solver for Step 2 (assuming it is solved to comparable and high-enough accuracy).

\subsection{Summary of the solver notation}
Three distinct ISA discretization methods are evaluated in this study: GISA,\cite{Verstraelen2012b} the minimum basis iterative stockholder analysis model (MBIS),\cite{Verstraelen2016} and LISA~\cite{Benda2022} for which we derived a variety of solvers in the previous sections~\ref{ssec:GlobalLagrangain} and \ref{ssec:AlternatingMinimization}.

For LISA-family methods, we introduce several notations to distinguish them.
On the one hand, we introduce two notations, \clsp{LISA} and \clsn{LISA}, to refer to all LISA solvers with and without the non-negative $c_{a,k}$ constraints, respectively.
On the other hand, depending on whether the global version of LISA is used, we employ the notations \clsg{LISA} (with prefix ``gLISA-'') and \clsa{LISA} (with prefix ``aLISA-'') to refer to the global and alternating versions of LISA solvers, respectively.
Here, ``global'' refers to the approach where a set of globally coupled equations is solved either by direct (global) minimization or by using a global solver for the root-finding problem (Section~\ref{ssec:GlobalLagrangain}), as opposed to the alternating minimization schemes (Section~\ref{ssec:AlternatingMinimization}).
Additionally, \clspg{LISA} is used to refer to all solvers that belong to the \clsp{LISA} and \clsg{LISA} categories, while \clsng{LISA} refers to all solvers in the \clsn{LISA} and \clsg{LISA} categories.
Similarly, the notations \clspa{LISA} (\clsna{LISA}) are used to refer to all solvers in the \clsp{LISA} (\clsn{LISA}) and \clsa{LISA} categories, respectively.
Table~\ref{tbl:lisa_categroy} summarizes all LISA sub-categories used in this work.
\begin{table}[htpb]
    \centering
    \caption{Summarized LISA sub-categories used in this work.
    }
    \begin{tabular*}{\textwidth}{@{\extracolsep{\fill}}cccc}
    \toprule
    &            & \multicolumn{2}{c}{LISA solvers}    \\
    \cmidrule{3-4}
    &            & \clsg{LISA}   & \clsa{LISA}   \\
    \midrule
    \multirow{2}{*}{LISA solvers}
    & \clsp{LISA} & \clspg{LISA}  & \clspa{LISA}  \\
    & \clsn{LISA} & \clsng{LISA}  & \clsna{LISA}  \\
    \bottomrule
    \end{tabular*}
    \label{tbl:lisa_categroy}
\end{table}

In theory, all \clsp{LISA} and \clsn{LISA} solvers should converge to the their respective optima owing to the uniqueness of the minimizer in convex optimization problems.
However, deviations exist between \clsg{LISA} and \clsa{LISA} solvers due to quadrature errors, which depend on the molecular grids employed. Generally, employing finer grids yields better results, though at an increased computational cost.
Ultimately, all solvers across the four sub-categories are expected to converge to their respective numerically unique solutions.

Moreover, the \clspa{LISA} solvers corresponds to the schemes introduced in the original contribution.\cite{Benda2022}
It should be noted that both GISA and MBIS methods can also be viewed as alternating optimization problem.
In the GISA scheme, the quadratic programming problem is solved, as noted in Ref.~\citenum{Benda2022}; therefore, it is denoted as \m{GISA-QUADPROG}.
For the MBIS scheme, due to the use of the self-consistent solver, it is denoted as \m{MBIS-SC} in this work.
All methods are listed in Table~\ref{tbl:isa_solvers}.
\begin{table}[htpb]
    \scriptsize
    \centering
    \caption{Summarized ISA solvers used in this work. See text for more details.
    }
    \label{tbl:isa_solvers}
    \begin{tabular*}{\textwidth}{@{\extracolsep{\fill}}cccc}
    \toprule
    Category-I & Category-II &  Notation                              & Algorithm \\
    \midrule
            & & \m{GISA-QUADPROG}                     & Eqs.~\eqref{eq:general-pos}, (45)-(46) in Ref.\citenum{Benda2022} \\
            & & \m{MBIS-SC}                           & Eqs.~\eqref{eq:general-pos}, and Ref.\citenum{Verstraelen2016} \\
              \midrule
     \multirow{4}{*}{\clsp{LISA}}        & \multirow{2}{*}{\clspg{LISA}} & \m{gLISA-CVXOPT}                      & Eqs.~\eqref{eq:general-pos},~\eqref{eq:alt-const-opt-pos},~\eqref{eq:common-constraint}-\eqref{eq:aim_weight_func},~\eqref{eq: gLISA_convex_obj_func}-\eqref{eq: gLISA_convex_hessian} \\
            & & \m{gLISA-SC}                          & Eqs.~\eqref{eq:general-pos},~\eqref{eq:alt-const-opt-pos},~\eqref{eq:common-constraint}-\eqref{eq:aim_weight_func},~\eqref{eq:fixed-point-global}-\eqref{eq:rho0_m} \\
            \cmidrule{2-4}
            &\multirow{2}{*}{\clspa{LISA}} & \m{aLISA-CVXOPT}                       & Eqs.~\eqref{eq:general-pos},~\eqref{eq:local-step2-pos},~\eqref{eq:lisa_convex_obj_func}-\eqref{eq:lisa_convex_hessian} \\
            & & \m{aLISA-SC}                           & Eqs.~\eqref{eq:general-pos},~\eqref{eq:local-step2-pos},~\eqref{eq:lisa_sc_eq}-\eqref{eq:lisa_sc_rho0_a} \\
              \midrule
    \multirow{10}{*}{\clsn{LISA}}        &\multirow{5}{*}{\clsng{LISA}} &\m{gLISA-FD-DIIS}                     & Eqs.~\eqref{eq:general},\eqref{eq:common-constraint}-\eqref{eq:aim_weight_func},~\eqref{eq:fixed-point-global}-\eqref{eq:rho0_m}, and Algorithm 2 in Ref.\citenum{Chupin2021} \\
            & &  \m{gLISA-R-DIIS}                      & Eqs.~\eqref{eq:general},~\eqref{eq:alt-const-opt}-\eqref{eq:aim_weight_func},~\eqref{eq:fixed-point-global}-\eqref{eq:rho0_m}, and Algorithm 3 in Ref.\citenum{Chupin2021} \\
            & &  \m{gLISA-AD-DIIS}                     & Eqs.~\eqref{eq:general},~\eqref{eq:alt-const-opt}-\eqref{eq:aim_weight_func},~\eqref{eq:fixed-point-global}-\eqref{eq:rho0_m}, and Algorithm 4 in Ref.\citenum{Chupin2021} \\
            & &  \m{gLISA-M-NEWTON}                    & Eqs.~\eqref{eq:general},~\eqref{eq:alt-const-opt}-\eqref{eq:aim_weight_func},~\eqref{eq:newton_h}-\eqref{eq:newton_hessian} with a line search \\
            & &  \m{gLISA-QUASI-NEWTON}                & Eqs.~\eqref{eq:general},~\eqref{eq:alt-const-opt}-\eqref{eq:aim_weight_func},~\eqref{eq:newton_h}-\eqref{eq:newton_hessian}, BFGS with a line search \\
            \cmidrule{2-4}
            &\multirow{5}{*}{\clsna{LISA}} &  \m{aLISA-FD-DIIS}                      & Eqs.~\eqref{eq:general},~\eqref{eq:local-step2},~\eqref{eq:lisa_sc_eq}-\eqref{eq:lisa_sc_rho0_a}, and and Algorithm 2 in Ref.\citenum{Chupin2021} \\
            & &  \m{aLISA-R-DIIS}                       & Eqs.~\eqref{eq:general},~\eqref{eq:local-step2},~\eqref{eq:lisa_sc_eq}-\eqref{eq:lisa_sc_rho0_a}, and Algorithm 3 in Ref.\citenum{Chupin2021} \\
            & &  \m{aLISA-AD-DIIS}                      & Eqs.~\eqref{eq:general},~\eqref{eq:local-step2},~\eqref{eq:lisa_sc_eq}-\eqref{eq:lisa_sc_rho0_a}, and Algorithm 4 in Ref.\citenum{Chupin2021} \\
            & &  \m{aLISA-M-NEWTON}                     & Eqs.~\eqref{eq:general},~\eqref{eq:local-step2},~\eqref{eq:merged_lagrangian}-\eqref{eq:lisa_newton_for_eq_jacobian} with a line search \\
            & &  \m{aLISA-QUASI-NEWTON}                 & Eqs.~\eqref{eq:general}, BFGS,~\eqref{eq:local-step2},~\eqref{eq:merged_lagrangian}-\eqref{eq:lisa_newton_for_eq_jacobian} with a line search \\
            \bottomrule
    \end{tabular*}
\end{table}

    \section{Computational details}
    \label{sec:computational-details}
    As anticipated in Section~\ref{sec:methods}, the implementation of the LISA-family methods requires numerical quadrature, and in this work, the Becke-Lebedev grids are employed for numerical integration over molecular volumes.\cite{Becke1988}
These grids consist of a set of atomic grids with atomic weight functions between neighboring atoms.
The atomic weight functions determine the contribution of each atom to the molecule, and in this work, Becke atomic weight functions are used.\cite{Becke1988}
Each atomic grid consists of radial and angular components.
For simplicity, 150 Gauss-Chebyshev radial points and 194 Lebedev-Laikov angular points were employed for all atoms, which is also the default setup in the \texttt{Denspart} package, where the global version of the \m{MBIS} method is implemented.\cite{Denspart2024}
The Gauss-Chebyshev integration interval $[-1, 1]$ is mapped into the semi-infinite radial interval $[0, +\infty)$ based on Ref.~\citenum{Becke1988}.\cite{Denspart2024}

The algorithms for GISA has been implemented according to the methodology outlined in Ref.\citenum{Benda2022}.
To solve quadratic problems (QP) in GISA, the ``qpsolver'' package was employed,\cite{qpsolvers2023} which offers a general interface for various QP solvers.
In this work, the ``quadprog'' solver was used.
The parameters for H, C, N, and O atoms were adopted from the previous work,\cite{Verstraelen2012b} whereas parameters for other elements were obtained using the procedure proposed in Ref.\citenum{Verstraelen2012b}.
The exponents for Gaussian functions are used for both GISA and LISA methods in this work, listed in Table S1, where a larger number of basis functions is used for heavier elements.
It should be noted that the results of some solver, e.g., \m{aLISA-SC} and \m{gLISA-SC}, depend on the initial values.
Specifically, initial values set to zero could consistently result in the corresponding $c_{a,k}$ being zero.
Therefore, the initial values are obtained by fitting only to the corresponding neutral atom density and are required to be positive and non-zero (with the lowest value being $10^{-4}$), as presented in Table S2.
Additionally, we scale all initial values to correspond with the molecular population, ensuring that the sum of all initial $c_{a,k}$ values equals the molecular population.
The implementation of the MBIS model is based upon in Ref.~\citenum{Verstraelen2016}.

The convergence criteria can vary among different solvers and Table~\ref{tbl:solver_options} lists the convergence criteria used in this work.
For all solvers employing the alternating minimization strategy, including \m{GISA-QUADPROG}, \m{MBIS-SC}, and \clsa{LISA} solvers, the outer iterations indexed by $m$ are stopped after the root-mean-square (rms) deviation (increment) $\epsOut$ between the pro-molecule densities of the last and the previous iterations drops below a threshold of $10^{-8}$:
\begin{align}
    \epsOut := \sqrt{\int_{\R^3}  [\rho^{0, (m, \ell_{\max})}(\br) - \rho^{0,(m-1, \ell_{\max})}(\br)]^2 \, d\br } < 10^{-8}.
    \label{eq:epsOut}
\end{align}
This criterion is also used for the \m{gLISA-SC}, \m{gLISA-M-NEWTON}, and \m{gLISA-QUASI-NEWTON} solvers.
The criterion chosen for inner iterations indexed by $\ell$ (if applicable) depends on the solver.
For the \m{GISA-QUADPROG} solver, the convergence threshold is related to machine precision as this comes with the used software package.

The \m{MBIS-SC}, \m{aLISA-SC}, \m{aLISA-M-NEWTON}, and \m{aLISA-QUASI-NEWTON} solvers are stopped after the rms deviation (increment) $\epsInn$ between the pro-atom densities of the last and the previous iteration drops below a threshold of $10^{-12}$ (see Ref.~\citenum{Verstraelen2012b}):
\begin{align}
    \epsInn := \max_{a=1,\ldots,M} \sqrt{\int_{\R^3}  [\rho_a^{0, (m, \ell)}(\br) - \rho_a^{0,(m, \ell-1)}(\br)]^2 \, d\br } < 10^{-12}.
    \label{eq:epsInn}
\end{align}

We use the same threshold for both \m{aLISA-CVXOPT} and \m{gLISA-CVXOPT} listed in Table~\ref{tbl:solver_options}, and the meaning of the convergence options can be found in Ref.\citenum{CVXOPT}.
The criteria of \m{aLISA-CVXOPT} used for the inner loop can reproduce the same results as the ones obtained with \m{aLISA-SC}.
Specifically, the deviation of the number of outer iterations between \m{aLISA-CVXOPT} and \m{aLISA-SC} are less than one.
The threshold chosen for DIIS is the $L_2$ norm of the residual error vector ($\epsilon_{\Vert r \Vert_2}$), set to $10^{-12}$ for all cases.
Additional input options for DIIS variants are provided in Table~\ref{tbl:solver_options}, with their explanations available in Ref.~\citenum{Chupin2021}.

The use of different thresholds in Eq.~\eqref{eq:epsOut} and Eq.~\eqref{eq:epsInn} is motivated by the structure of the nested iterative procedures involved in our calculations.
Specifically, these procedures consist of an outer and an inner iteration.
To ensure that numerical errors introduced during the inner iteration do not propagate into the outer iteration, we set a more stringent tolerance for the inner loop.
In practice, the tolerance for the inner iteration is chosen to be $10^4$ smaller than that of the outer iteration.
This difference ensures that any inaccuracies in the inner loop remain well below the threshold that could affect the convergence of the outer loop.
Although the inner tolerance can be relaxed in certain applications without influencing the number of outer iterations, maintaining a conservative threshold helps guarantee overall numerical stability.
This approach provides robust convergence while ensuring that the solution retains high accuracy throughout the iterative process.

\begin{table}[htpb]
\scriptsize
\centering
\caption{Summarized convergence criteria used in all solvers.}
\begin{tabular*}{\textwidth}{@{\extracolsep{\fill}}ccccc}
    \toprule
    Categroy-I & Categroy-II & Solver              & Outer loop options                    & Inner loop options  \\
    \midrule
    & & \m{GISA-QUADPROG}       &$\epsOut < 10^{-8}$    & Machine precision    \\
    & & \m{MBIS-SC}           &$\epsOut < 10^{-8}$    & $\epsInn < 10^{-12}$  \\
    \midrule
    \multirow{4}{*}{\clsp{LISA}}& \multirow{2}{*}{\clspg{LISA}} & \m{gLISA-CVXOPT}        & $\epsilon_\text{feas}<10^{-8}$, $\epsilon_\text{abs}<10^{-7}$, and $\epsilon_\text{rel}<10^{-6}$  & $--$ \\
    & & \m{gLISA-SC}            & $\epsOut < 10^{-8}$   & $--$        \\
    \cmidrule{2-5}
    &\multirow{2}{*}{\clspa{LISA}}  & \m{aLISA-CVXOPT}         & $\epsOut < 10^{-8}$   & $\epsilon_\text{feas}<10^{-8}$, $\epsilon_\text{abs}<10^{-7}$, and $\epsilon_\text{rel}<10^{-6}$     \\
    & & \m{aLISA-SC}             & $\epsOut <10^{-8}$    & $\epsInn < 10^{-12}$ \\
    \midrule
   \multirow{10}{*}{\clsn{LISA}}& \multirow{5}{*}{\clsng{LISA}} & \m{gLISA-FD-DIIS}       & $\epsilon_{\Vert r \Vert_2} < 10^{-8}$, $q=8$                           & $--$ \\
    & & \m{gLISA-R-DIIS}        & $\epsilon_{\Vert r \Vert_2} < 10^{-8}$, $\tau<10^{-3}$                  & $--$  \\
    & & \m{gLISA-AD-DIIS}       & $\epsilon_{\Vert r \Vert_2} < 10^{-8}$, $\delta<10^{-4}$                & $--$ \\
    & & \m{gLISA-M-NEWTON}      & $\epsOut < 10^{-8}$   & $--$       \\
    & & \m{gLISA-QUASI-NEWTON}  & $\epsOut < 10^{-8}$   & $--$  \\
    \cmidrule{2-5}
    &\multirow{5}{*}{\clsna{LISA}}  & \m{aLISA-M-NEWTON}       & $\epsOut < 10^{-8}$   & $\epsInn < 10^{-12}$ \\
    & & \m{aLISA-QUASI-NEWTON}   & $\epsOut < 10^{-8}$   & $\epsInn < 10^{-12}$\\
    & & \m{aLISA-R-DIIS}         & $\epsOut < 10^{-8}$   &  $\epsilon_{\Vert r \Vert_2} < 10^{-12}$, $\tau<10^{-2}$ \\
    & & \m{aLISA-FD-DIIS}        & $\epsOut < 10^{-8}$   &  $\epsilon_{\Vert r \Vert_2} < 10^{-12}$, $q=5$  \\
    & & \m{aLISA-AD-DIIS}        & $\epsOut < 10^{-8}$   &  $\epsilon_{\Vert r \Vert_2} < 10^{-12}$, $\delta<10^{-2}$\\
    \bottomrule
\end{tabular*}
\label{tbl:solver_options}
\end{table}

All calculations were performed using the \texttt{Horton-Part} module~\cite{YingXing2023} on a MacBook Pro equipped with an Apple M2 Pro chip, featuring 12 cores (8 performance cores and 4 efficiency cores) and 16 GB of memory.
The development of \texttt{Horton-Part} is based on the \texttt{part} submodule from the \texttt{Horton} package.\cite{Chan2024}
Since version 1.0.0 of \texttt{Horton-Part}, the new \texttt{Grid} package has been utilized, replacing the older version in \texttt{Horton}.\cite{Tehrani2024}
The \texttt{IOData}~\cite{Verstraelen2021} and \texttt{GBasis}~\cite{Kim2024} packages were employed to prepare the molecular density and its gradients on grid points.

For benchmark testing in this study, we utilized 42 organic and inorganic molecules from the TS42 dataset,\cite{Cheng2022} along with additional six charged molecular ions and anions from Ref.~\citenum{Heidar-Zadeh2018}, where the molecular structures were optimized using DFT at the B3LYP/aug-cc-pVDZ level with GAUSSIAN16.\cite{g16}
The LDA/aug-cc-pVDZ level of theory~\cite{Kohn1965, Kendall1992} was used for molecular density calculations because it resulted in a good correspondence with experimental data in previous work.\cite{Cheng2022}
It should be noted that benchmarking different levels is beyond the scope of this work and will be investigated in future work.

    \section{Results and discussion}
    \label{sec:res}
    \subsection{Properties}
\label{ssec:properties}
Table~\ref{tbl:solver_conv} summarizes the convergence performance of various computational solvers on 48 molecules, by detailing the number of molecules for which specific optimization constraints were met.
Specifically, $N(c_{a,k})$ denotes the number of molecules where the optimized parameter $c_{a,k}$ fell below $-10^{-4}$, corresponding to the negative $c_{a,k}$.
We observed convergence issues during the optimization process if the pro-atom density, i.e., $\rho_a^0$, dropped below $-10^{-12}$.
Therefore, the column labeled $N_\text{Conv.}$ shows the total number of molecules for which the solver successfully converged.
The specific molecules for which the solvers did not converge are listed as well.
The numerical observations confirm the expected behavior as stated after the definition of each method in Section~\ref{sec:methods}.
Specifically, solvers like \m{GISA-QUADPROG}, \m{MBIS-SC}, and all \clsp{LISA} solvers maintain non-negative values of both $c_{a,k}$ and $\rho_a^{0}$ when non-negative initial values are provided.
Not all \clsn{LISA} solvers converge for all molecules; for example, \m{aLISA-R-CDIIS}, \m{aLISA-AD-CDIIS}, \m{aLISA-FD-CDIIS}, \m{gLISA-R-CDIIS}, \m{gLISA-AD-CDIIS}, and \m{gLISA-QUASI-NEWTON} do not.
However, this does not imply that they cannot converge with any user-tailored parameters.

We found that there are of 38 molecules where the optimized $c_{a,k}$ values include negative numbers for \clsn{LISA} solvers, such as \m{gLISA-M-NEWTON}, \m{aLISA-M-NEWTON}, and \m{aLISA-QUASI-NEWTON}.
The \m{gLISA-FD-DIIS} solver converges for all test molecules; however, it does not always yield the expected correct negative $c_{a,k}$ values, in comparison to \m{gLISA-M-NEWTON}.
It can also be observed that the \m{gLISA-QUASI-NEWTON} solver converges for all neutral molecules and negatively charged molecules but not for positively charged ones.
Additionally, it still achieves non-negative values for $c_{a,k}$ for \ce{SO2} and \ce{SiH4}, whereas negative $c_{a,k}$ values are obtained using the \m{gLISA-M-NEWTON} solver.
This can be attributed to the use of nearly linearly-dependent basis functions due to similar $\alpha_{a,k}$ coefficients for \ce{Si} (9.51 and 7.87) and \ce{S} (17.64 and 17.51) atoms as shown in Table S1.

For simplicity, in the subsequent discussion, we analyze only those solvers that consistently yield converged results, as indicated by $N_\text{Conv.}=48$ in Table~\ref{tbl:solver_conv}.

\InputIfFileExists{convergence.itex}{}{%
    \message{LaTeX Warning: File `%
    \filename@base.\ifx\filename@ext\relax tex\else\filename@ext\fi'%
    not found on input line \the\inputlineno}

%
%
\subsection{Performance}
\label{ssec:performance}
Figure~\ref{fig:benchmark} (a) displays a comparison of all ISA schemes with different solvers that yield converged results in terms of the number of iterations in the outer iterations (indexed by $m$).
\m{gLISA-FD-DIIS} results are not presented because it failed to find all the required negative optimized $c_{a,k}$ values compared to \m{gLISA-M-NEWTON} discussed in Section~\ref{ssec:properties}.
We also compare the total time used for the molecular density partitioning, as shown in Figure~\ref{fig:benchmark} (b).
While the time usage for outer iterations remains consistent across alternating methods, variation in time spent on inner iterations arises from differing characteristics among solvers.
Thus, in absolute timings the variation among all \clsa{LISA} solvers is small since the Step 2 calculations are all local and thus independent problems.
In consequence, the time spent in Step 2 is relatively small compared to the time spent in Step 1 which is a global update.

Several observations can be made:
\begin{enumerate}
    \item
        The number of outer iterations in \m{MBIS-SC} is lower than in all \clsa{LISA} methods but higher than all \clsg{LISA} methods, except for the \m{gLISA-SC} method.
        Additionally, it is slightly higher than in the \m{GISA-QUADPROG} method.
        This leads to the conclusion that the \m{MBIS-SC} method is faster than all \clsa{LISA} methods, as shown in Figure~\ref{fig:benchmark} (b), because recalculating AIM weights in Step 1 (Eq.~\eqref{eq:step1}) is the main computational cost.
        This is also the reason why all \clsa{LISA} methods, except for \m{aLISA-CVXOPT} where the Hessian matrix is required and computationally expensive, have very similar total time usage.
    \item
        These numerical results confirm the theory that all \clspa{LISA} or \clsna{LISA} methods converge to their unique solution~\eqref{eq:general-pos} or~\eqref{eq:general} respectively using the same number of outer iterations when a similar convergence criterion for inner iterations is applied (with the exception of \ce{H3O+} where the solution lies extremely close to the boundary of the feasible set).
    \item
        The number of outer iterations of the \m{gLISA-SC} solver is much higher than for all other solvers, as shown in Figure~\ref{fig:benchmark} (a), resulting to it being the most computationally costly solver.
        In contrast, the \m{gLISA-CVXOPT} solver has nearly the lowest number of outer iterations among all solvers, but it still incurs a higher computational cost for large molecules than other solvers, except for the \m{gLISA-SC} solver.
        This can be attributed to the requirement for gradient and Hessian matrices during the optimization, which are normally computationally expensive.
    \item
        Although the \m{gLISA-M-NEWTON} method has the lowest number of outer iterations, its time usage is higher for larger molecules compared to the \m{MBIS-SC} method due to the costly calculations of the Jacobian matrix.
        However, this can be improved using the quasi-Newton method, i.e., the \m{gLISA-QUASI-NEWTON} method.
        The total number of outer iterations increases, but it is still lower than that of the \m{MBIS-SC} method, as shown in Figure~\ref{fig:benchmark} (a).
        Besides the approximation of the Jacobian matrix, there is no matrix inversion in the quasi-Newton method.
        Therefore, the total time usage of the \m{gLISA-QUASI-NEWTON} method significantly reduces.
        Some exceptions are observed only for diatomic molecule systems, where \m{gLISA-M-NEWTON} and \m{MBIS-SC} are slightly faster.
        One potential issue with the \m{gLISA-QUASI-NEWTON} method is that the line search might fail, i.e., $\rho_{a}(\mathbf{r}) < 0$ for some $\mathbf{r}$.
        For instance, in the case of a positively charged molecule ions, e.g., \ce{CH3+} and \ce{H3O+},
        has shown signs of non-robustness while performing remarkably well on neutral or negatively charged molecules, which has been discussed in Section~\ref{ssec:properties}.
\end{enumerate}
\begin{figure}[h]
    \begin{center}
        \includegraphics{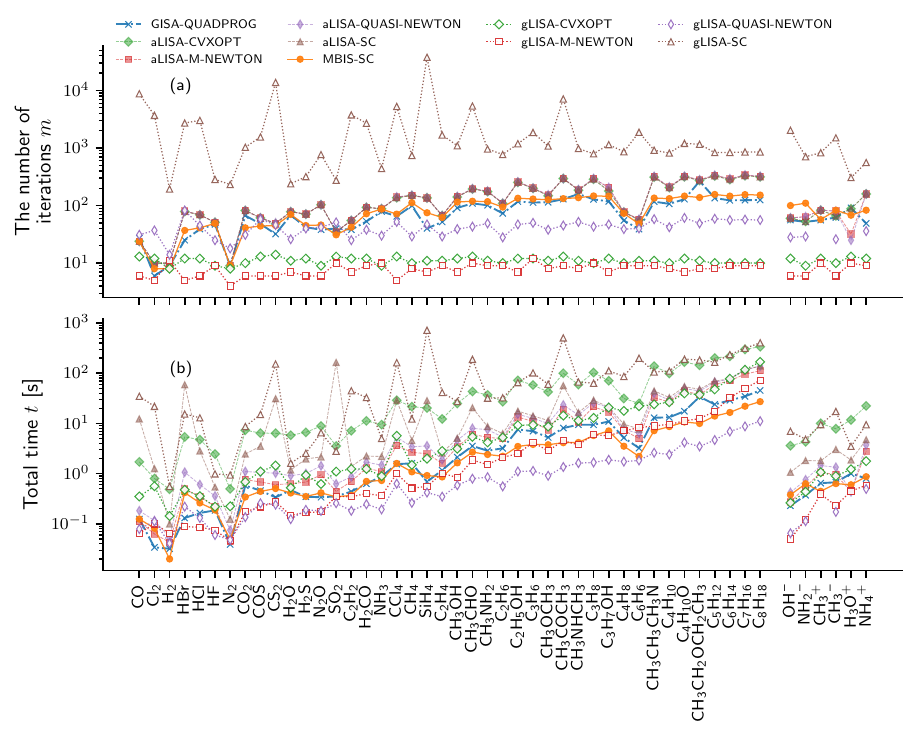}
    \end{center}
    \caption{
        Benchmark comparison of the GISA, MBIS, and LISA partitioning schemes tested on a set of 48 molecules.
        Results for the LISA method are shown for various solvers, as defined in the main text.
        (a) Number of outer iterations required to achieve convergence.
        (b) Total computational time needed for molecular density partitioning to reach convergence.
    }
    \label{fig:benchmark}
\end{figure}
%
%
\subsection{Convergence and accuracy}
In this section, we first investigate the consistency and precision of entropy calculations within the LISA framework, specifically examining the effects of negative (from \clsn{LISA} solvers) and non-negative (from \clsp{LISA} solvers) parameters on the entropy outcomes.
All entropies are evaluated using Eq.~\eqref{eq: gLISA_convex_obj_func} with molecular grids.
For molecules with non-negative optimized $c_{a,k}$ values, the entropy differences are observed to be less than $10^{-5}$ between \clspa{LISA} and \clsna{LISA} solvers.
In cases where the pro-atom $c_{a,k}$ coefficients can take negative, \clsna{LISA} solvers are shown to achieve lower entropy values compared to \clspa{LISA} solvers, with the difference maintained below $10^{-4}$ except for \ce{H3O+} where the difference is $-2.5\times 10^{-3}$.
A detailed comparison of the converged entropy between \clsna{LISA} (represented by \m{aLISA-M-NEWTON}) and \clspa{LISA} (represented by \m{aLISA-SC}) solvers can be found in Table~\ref{tbl:entropy_newton_vs_sc}.
The relative error, $\delta$, is computed as $\delta = (s_2 - s_1) / s_1 \times 100$, where $s_1$ and $s_2$ are the entropies obtained from the \m{aLISA-M-NEWTON} and \m{aLISA-SC} solvers, respectively.
In principle, $\delta \ge 0$ should always hold, but this is not the case for the \ce{SO2} molecule.
This inconsistency arises due to differences in the quadrature schemes used for parameter optimization and entropy calculations for \clsa{LISA} solvers.
Specifically, in the parameter optimization of \clsa{LISA} methods, atomic grids are employed, whereas the entropy, defined in Eq.~\eqref{eq: gLISA_convex_obj_func}, is evaluated on molecular grids to maintain consistency with the \clsg{LISA} methods.
We also evaluated the entropy using Eq.~\eqref{eq:entropy} with atomic grids for \ce{SO2}, yielding a value of 0.05622956 from the \m{aLISA-M-NEWTON} solver, as expected, which is less than the 0.05623555 obtained from the \m{aLISA-SC} solver.
In addition, the entropy difference between any \clsg{LISA} solvers and \clspa{LISA} is less than $6 \times 10^{-3}$.
\InputIfFileExists{last_entropy.itex}{}{%
    \message{LaTeX Warning: File `%
    \filename@base.\ifx\filename@ext\relax tex\else\filename@ext\fi'%
    not found on input line \the\inputlineno}

Figure~\ref{fig:last_entropy} presents the results of the converged entropies obtained using different solvers.
In general, \m{MBIS-SC} converges to the highest entropy, followed by the \m{GISA-QUADPROG} solver, while all LISA-family methods converge to the lowest entropy.
Two exceptions are observed, i.e., \ce{NH2-} and \ce{CH3-}, where \m{MBIS-SC} has a lower entropy compared to both \m{GISA-QUADPROG} and any LISA-family solvers.
This can be attributed to the lack of more diffused basis functions in GISA or LISA, because the Gaussian exponential coefficients are obtained by fitting ions with only 1 a.u. charge.
By simply adding an extra Gaussian (Slater) basis function with an exponential coefficient of $0.1$ ($1.0$) for the H atom, a lower entropy can be obtained for \ce{CH3-}, \ce{NH2-} and \ce{OH-} using \m{aLISA-M-NEWTON} or \m{aLISA-SC} as shown in Table~\ref{tbl:charged_entropy}.

\begin{figure}[h]
    \begin{center}
        \includegraphics{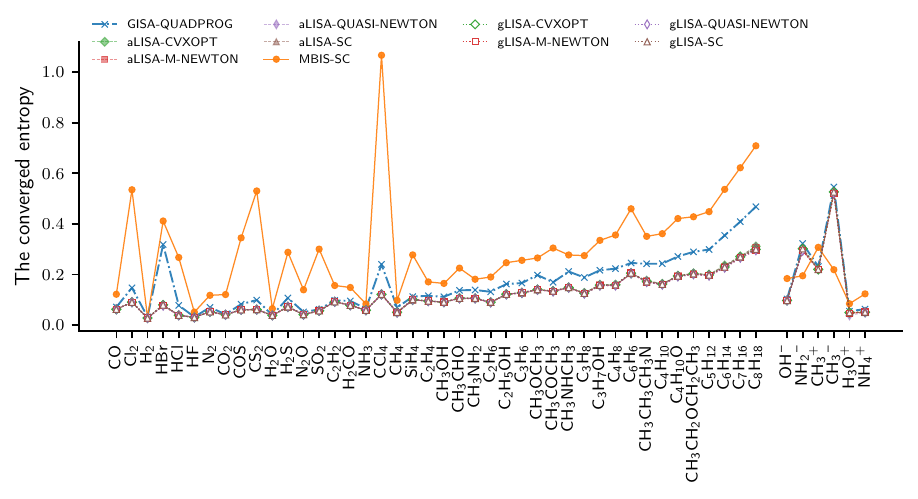}
    \end{center}
    \caption{
        Benchmark comparison of the GISA, MBIS, and LISA partitioning schemes tested on a set of 48 molecules.
    Comparison of the entropy values at convergence.
    Results for the LISA method are presented for various solvers, as defined in the main text.
}
    \label{fig:last_entropy}
\end{figure}

\begin{table}[h!]
\footnotesize
\centering
\caption{
Converged entropy comparison between \m{MBIS-SC} and \m{aLISA-SC} (\m{aLISA-M-NEWTON})  with the addition of an extra Gaussian (Slater) basis function, each with exponential coefficients equal to $0.1$ ($1.0$).
\clspa{LISA}@1g (\clsna{LISA}@1g) and \clspa{LISA}@1s (\clsna{LISA}@1s) denote \m{aLISA-SC} (\m{aLISA-M-NEWTON}) with the extra Gaussian and Slater basis functions, respectively.
}
\begin{tabular*}{\textwidth}{@{\extracolsep{\fill}}cccccc}
    \toprule
    Molecule             & \m{MBIS-SC} & \clspa{LISA}@1g  & \clsna{LISA}@1g & \clspa{LISA}@1s & \clsna{LISA}@1s    \\
    \midrule
    \ce{CH3-}            & $0.2185$ & $0.1905$      & $0.1752$    & $0.1763$     & $0.1614$       \\
    \ce{OH-}             & $0.1834$ & $0.0479$      & $0.0453$    & $0.0471$     & $0.0444$       \\
    \ce{NH2-}            & $0.1943$ & $0.1098$      & $0.1056$    & $0.1112$     & $0.1071$       \\
    \bottomrule
\end{tabular*}
\label{tbl:charged_entropy}
\end{table}

Next, we compare the convergence behavior of the entropy for different ISA methods.
Figure~\ref{fig:entropy} displays the entropies obtained by different ISA methods at each outer iteration for four example molecules, i.e., \ce{C2H2}, \ce{C2H4}, \ce{C2H5OH}, and \ce{C2H6}.
The results of other molecules can be found in the Supporting Information.
To facilitate a clear comparison, we only consider the first 15 iterations for all solvers, and all iterations are included for solvers where the total number of outer iterations is less than 15, e.g., \m{gLISA-CVXOPT} and \m{gLISA-M-NEWTON}\@.
As discussed in Ref.~\citenum{Benda2022}, all variants of \clsa{LISA} with non-negative $c_{a,k}$ (\clspa{LISA}) or $c_{a,k}$ with both signs (\clsna{LISA}) should maintain consistency in the number of iterations, entropies, atomic charges, and $c_{a,k}$ values at each outer iteration, due to the unique solution of the strictly convex optimization problem.
Therefore, we consider only the results of \m{aLISA-SC} and \m{aLISA-M-NEWTON} as specific instances representing \clspa{LISA} and \clsna{LISA} solvers.
As mathematically proven in Ref.~\citenum{Benda2022}, the entropy of the \clsa{LISA} solvers decays monotonically.
All \clsa{LISA} methods converge slightly faster than \m{GLISA-SC} but generally slower than other solvers, given the initial values used in this work, as shown in Figures~\ref{fig:benchmark}(a) and~\ref{fig:entropy}.
Additionally, our numerical tests demonstrate that among all \clsg{LISA} methods only \m{gLISA-SC} and \m{gLISA-CVXOPT} possess the characteristic of monotonically decreasing entropy.
\m{gLISA-SC} exhibits slower convergence compared to all other LISA-family methods.
\m{gLISA-CVXOPT} converges faster than all other LISA methods except for \m{gLISA-M-NEWTON}, but the issues of computational efficiency may impede its practical applications.
The entropy of the second iteration of \m{gLISA-M-NEWTON} is higher than that of other LISA-family solvers, except for \m{gLISA-CVXOPT}, because the identity matrix is always used as the initial Hessian matrix.
For \m{gLISA-CVXOPT}, the $c_{a,k}$ is optimized with respect to inequality constraints, which leads to slower entropy convergence in the first few iterations.

\begin{figure}[h!]
    \centering
    \includegraphics[scale=1.0]{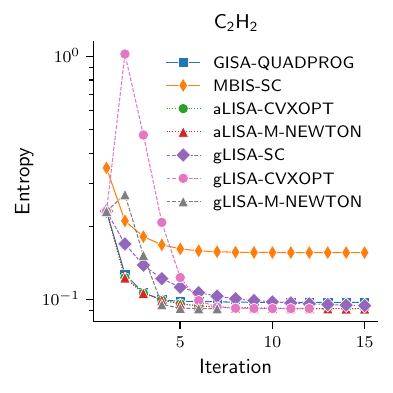}
    \includegraphics[scale=1.0]{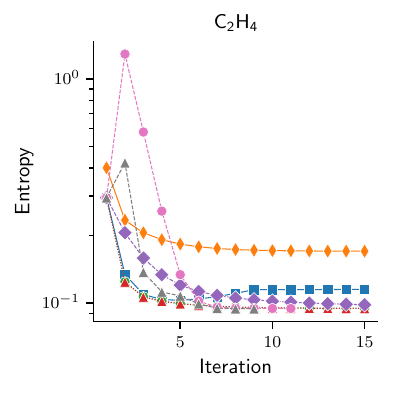} \\
    \includegraphics[scale=1.0]{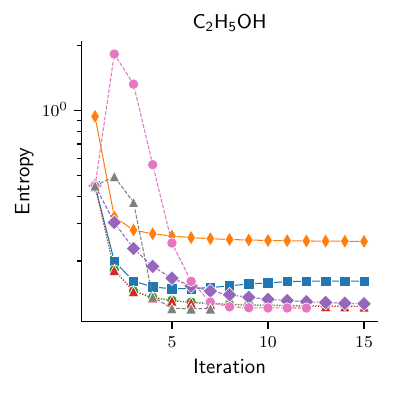}
    \includegraphics[scale=1.0]{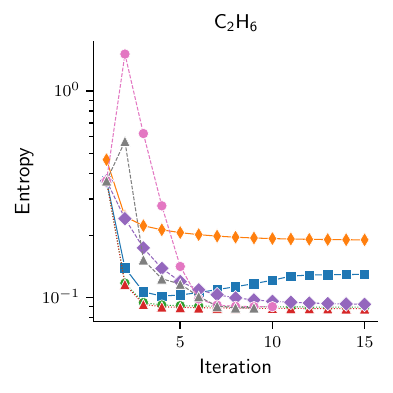}
    \caption{
        Comparison of the converged entropy for the GISA, MBIS, and LISA partitioning schemes tested on \ce{C2H2}, \ce{C2H4}, \ce{C2H5OH}, and \ce{C2H6}.
        Results for the LISA method with various solvers, as defined in the main text, are presented.
}
    \label{fig:entropy}
\end{figure}
%
%
\subsection{Comparison of AIM charges}
First, we observed that all \clspa{LISA}, \clsna{LISA}, \clsng{LISA}, and \clspg{LISA} solvers generally converge to their respective unique solutions as expected, and the corresponding AIM charges are reasonably close to each other, with the maximum charge difference being less than 0.01 a.u.\@
However, a few exceptions include the
\clsna{LISA} solvers (\m{aLISA-M-NEWTON} and \m{aLISA-QUASI-NEWTON}) for \ce{CH3-}, \ce{H3O+}, and \ce{NH2-},
and the \clsng{LISA} (\m{gLISA-M-NEWTON}) solver for \ce{CH3-} and \ce{NH2-}
which, compared to \clspa{LISA}, yield results that differ between 0.01 and 0.09.
This suggests that allowing negative $c_{a,k}$, i.e., \clsn{LISA}, could play an important role in systems of charged molecules.
The difference in the comparison between all \clsg{LISA} and \clsa{LISA} methods is less than 0.007, except for \ce{CCl4} and \ce{H3O+}, where a difference of 0.011 is observed.
It should be noted that for \ce{H3O+}, the larger difference is only observed in the comparison between the \clsng{LISA} and \clsna{LISA} methods.
For simplicity and clarity, in the following analysis, we will consider AIM charges obtained by the \clsa{LISA} solvers, and use \m{aLISA-M-NEWTON} and \m{aLISA-SC} solvers as representative of \clsna{LISA} and \clspa{LISA} solvers, respectively.
For simplicity, we use GISA, MBIS, \clsna{LISA}, and \clspa{LISA} to refer to \m{GISA-QUADPROG}, \m{MBIS-SC}, \m{aLISA-M-NEWTON}, and \m{aLISA-SC}, respectively.

Figure~\ref{fig:charges} shows a series of scatter plots comparing AIM charges via the GISA, MBIS, and \clsna{LISA} methods on 48 test molecules.
It should be noted that the \clspa{LISA} are not included due to the tiny difference between \clsna{LISA} and \clspa{LISA} in AIM charges.
A consistent linear correlation exists across all elements, signifying a robust relationship between the GISA and MBIS methods respectively and the \clsna{LISA} AIM charges.
This uniformity underscores the consistency in computational approaches for AIM charge determination.
Most data points of MBIS for the majority of elements closely follow the diagonal line (represented as a dashed line), indicating strong methodological agreement between MBIS and \clsna{LISA}.
However, a few outliers suggest possible variations in charge calculations.
\begin{figure}[h!]
    \begin{center}
        \includegraphics{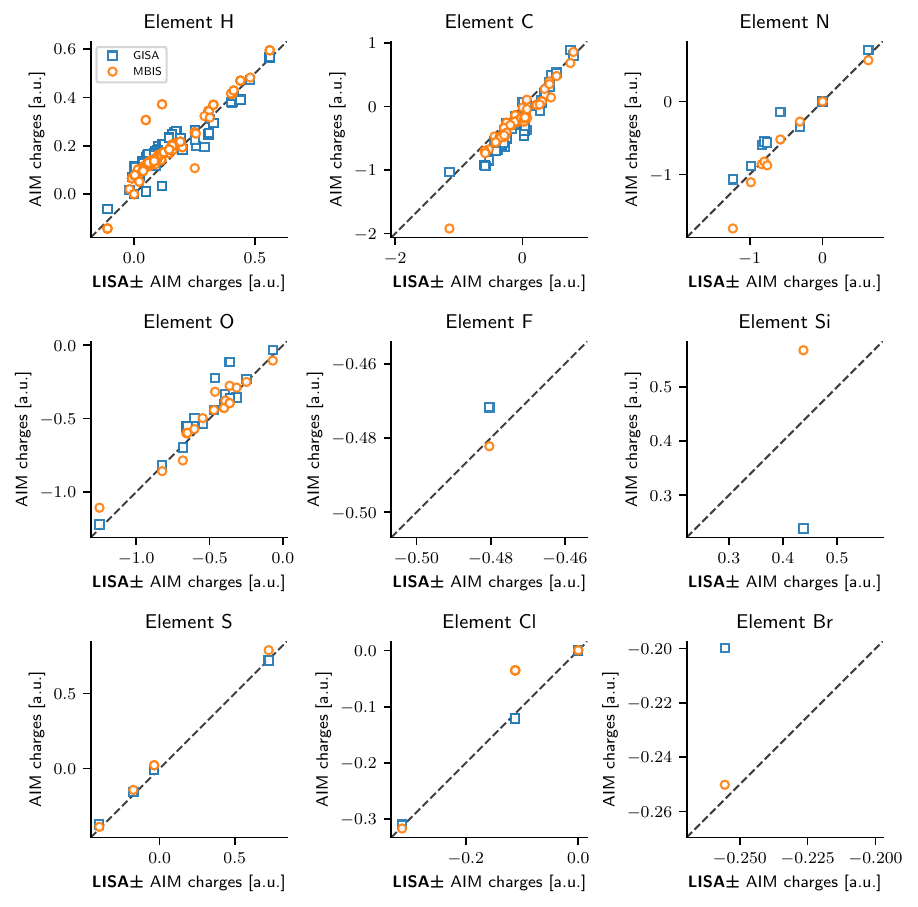}
    \end{center}
    \caption{Scatter plots comparing AIM charges (in a.u.) for nine elements across 48 test molecules, obtained using the GISA, MBIS, and \clsna{LISA} methods.
    The \clspa{LISA} results are omitted due to negligible differences compared to \clsna{LISA}.
    A dashed line representing perfect correlation ($y = x$) is included to illustrate the agreement between methods.
}
    \label{fig:charges}
\end{figure}

To study the deviations in AIM charges due to different methods, we compared AIM charges among the GISA, MBIS, and \clsna{LISA}, methods for \ce{CCl4}, \ce{CS2}, and \ce{SiH4} in the TS42 dataset and all charged molecules from the Ref.~\citenum{Heidar-Zadeh2018}.
The outliers in the plots are primarily attributed to these molecules.
The results are listed in Table~\ref{tbl:aim_charges}.
Some computational values available in the literature are also listed for comparison.
For the \ce{CCl4} and \ce{CS2} molecules, the electrostatic potential (ESP) fitting charges are included.\cite{Torii2003,Torii2004}
For all charged molecules, we compiled the AIM values obtained using the Hirshfeld-E (HE) method at the UB3LYP/aug-cc-pVDZ level from Ref.~\citenum{Heidar-Zadeh2018}.

Several key observations arise from the comparison between the \clsa{LISA} and MBIS methods for the \ce{CCl4} and \ce{CS2} molecules.
For \ce{CCl4}, the charge on the carbon atom calculated using the \clsna{LISA} (\clspa{LISA}) method is 0.450 (0.448), which aligns more closely with the ESP charges of 0.422 (Ref.~\citenum{Torii2003}) and 0.380 (Ref.~\citenum{Torii2004}) obtained by fitting both atomic charges and quadrupole moments, compared to the MBIS estimate of 0.140.

For \ce{CS2}, the \clsna{LISA}/\clspa{LISA} methods yield a carbon atom charge of 0.073, closer to the ESP value of 0.088,\cite{Torii2003} obtained by fitting both atomic charges and quadrupole moments, where the molecular quadrupole moment is properly reproduced.
In contrast, the MBIS method predicts a negative charge of $-0.044$.
We also evaluated the traceless molecular quadrupole moment, as defined by Buckingham,\cite{Buckingham1967} for \ce{CS2} using AIM densities from the \clsna{LISA} and MBIS methods.
Based on the quadrupole moment analysis of the \ce{CS2} molecule, both the \clsna{LISA} and MBIS methods reproduce the quadrupole moment relative to the reference value, although the signs of the atomic charges differ between the two methods.
The contribution of the atomic quadrupole moment of the C atom to the $zz$ component of the molecular quadrupole moment ($Q_{zz}$) obtained by MBIS ($0.255$ a.u.) and \clsna{LISA} ($0.397$ a.u.) is slightly lower than the value obtained from the ESP method (0.537 a.u.), as the dipole contribution is neglected in the ESP fitting.\cite{Torii2003}

Interestingly, the AIM charges obtained by \clsna{LISA} and \clspa{LISA}, through the addition of an extra Gaussian (Slater) basis function with exponential coefficients equal to 0.1 (1.0) for the H atom, are also listed in the table.
The differences between the \clsna{LISA} and \clspa{LISA} methods are less than 1\% except for \ce{CH3-} for which the differences are 7\%.

Furthermore, GISA shows a trend similar to \clsa{LISA} for both \ce{CCl4} and \ce{CS2} molecules.
However, the AIM charges produced by GISA for \ce{CS2} are smaller than those from \clsa{LISA}.
Notably, for \ce{SiH4}, the AIM charges from GISA are significantly lower compared to those from \clsa{LISA} and MBIS.

It has been observed that MBIS can yield anomalously negative values for anionic molecules.\cite{Heidar-Zadeh2018}
For all charged molecules, the \clsa{LISA} results demonstrate good agreement with the HE values computed in Ref.~\citenum{Heidar-Zadeh2018}.
In contrast, large discrepancies are evident between the MBIS and HE results for \ce{OH-} and \ce{NH2-}, consistent with findings from previous work.\cite{Heidar-Zadeh2018}
It should be noted that the difference of AIM charges obtained by \clsa{LISA} with and without extra basis functions is very small.
This implies that even when \clsa{LISA} converge to a higher entropy compared to MBIS, the former solvers can still predict reasonable AIM charges.
The extra diffuse basis function seems to mainly contribute to the tail of the atomic density.
However, this could be more important in force field development and in calculations of higher-ranking atomic moments and polarizabilities.\cite{Misquitta2014b,Verstraelen2016}
\begin{table}[h!]
\footnotesize
\centering
\caption{Comparison of atomic charges among the \m{GISA-QUADPROG} (GISA), \m{aLISA-M-NEWTON} (\clsna{LISA}), \m{aLISA-SC} (\clspa{LISA}) and \m{MBIS-SC} (MBIS) methods for molecules \ce{CCl4} and \ce{CS2}, along with available values from the literature.
The values in the parenthesis for the \clsna{LISA}/\clspa{LISA} solver are computed by adding an extra Gaussian and Slater basis with exponential coefficients equal to $0.1$ and $1.0$, respectively.
}
\begin{tabular*}{\textwidth}{@{\extracolsep{\fill}}ccrrrrc}
    \toprule
    Molecule            & Atom   & GISA     & \clsna{LISA}           & \clspa{LISA}  & MBIS   & Others                    \\
    \midrule
    \ce{CCl4}           & C      & $ 0.485$     & $0.450$            & $0.448$  & $ 0.140$  & $0.422$ (Ref.~\citenum{Torii2003}) and $ 0.380$ (Ref.~\citenum{Torii2004})   \\
    \ce{CS2}            & C      & $ 0.020$     & $0.073$            & $0.073$  & $-0.044$  & $0.088$ (Ref.~\citenum{Torii2003})  \\
    \ce{SiH4}           & Si     & $0.239$      & $0.443$            & $0.443$  & $0.568$   &  $--$                                    \\
    \ce{CH3+}           & C      & $0.307 $     & $0.425$            & $0.428$  & $0.349 $  & $0.447$ (Ref.~\citenum{Heidar-Zadeh2018})                                     \\
    \ce{CH3-}           & C      & $-1.032$     & $-1.147$                      & $-1.067$                     & $-1.920$  & $-1.047$ (Ref.~\citenum{Heidar-Zadeh2018})                                     \\
                        &        &              &          ($-1.133$, $-1.146$)&          ($-1.039$, $-1.052$)&           &                                                                                \\
    \ce{H3O+}           & O      & $-0.698$     & $-0.683$ & $-0.693$           & $-0.785$  & $-0.731$ (Ref.~\citenum{Heidar-Zadeh2018})                                     \\
    \ce{OH-}            & O      & $-1.223$     & $-1.250$                      & $-1.250$                     & $-1.107$  & $-1.200$ (Ref.~\citenum{Heidar-Zadeh2018})                                     \\
                        &        &              &          ($-1.199$, $-1.205$)&          ($-1.181$, $-1.187$)&           &                                                                                \\
    \ce{NH4+}           & N      & $-0.565$     & $-0.761$ & $-0.761$           & $-0.877$  & $-0.871$ (Ref.~\citenum{Heidar-Zadeh2018})                                     \\
    \ce{NH2-}           & N      & $-1.069$     & $-1.183$                      & $-1.183$                     & $-1.743$  & $-1.244$ (Ref.~\citenum{Heidar-Zadeh2018})                                     \\
                        &        &              &          ($-1.190$, $-1.199$) &          ($-1.160$, $-1.173$)&           &                                                                                \\
    \bottomrule
\end{tabular*}
\label{tbl:aim_charges}
\end{table}


    \section{Conclusions}
    \label{sec:summary}
    In this study, we conducted a comprehensive numerical analysis of various ISA solvers employed in molecular density partitioning, including MBIS, GISA, and the recently developed LISA schemes.
Initially, we adapted the original LISA approach to a broader framework by removing the non-negativity constraint on the pro-atomic expansion coefficients, $c_{a,k}$.
This modification resulted in two subcategories of LISA variants, denoted as \clsp{LISA} and \clsn{LISA}, which correspond to LISA methods with and without the non-negativity requirement on $c_{a,k}$, respectively.
Next, we derived an equivalent global version of LISA, denoted \clsg{LISA}, in contrast to the previously used two-step alternating \clsa{LISA} scheme.
Based on this, LISA-family methods can be classified into two categories, denoted as \clsg{LISA} and \clsa{LISA}, with the prefixes ``gLISA-'' and ``aLISA-'' indicating the global and alternating algorithm versions, respectively.

By examining the critical points of the Lagrangian associated with either the global or local constrained convex optimization problem, we formulated the problem as a set of nonlinear equations.
This alternative approach provides two equivalent formulations for solving the original convex optimization problem as either a root-finding problem or a fixed-point problem.
Thus, alternative optimization algorithms can be employed for both \clsa{LISA} and \clsg{LISA}.
For the root-finding problem, we used a Newton solver, while for the fixed-point problem, we utilized an iterative self-consistent solver, along with acceleration techniques such as DIIS.
Combining \clsp{LISA}/\clsn{LISA} with \clsg{LISA}/\clsa{LISA} produces four distinct subcategories of LISA variants, i.e., \clspg{LISA}, \clspa{LISA}, \clsng{LISA}, and \clsna{LISA}, each converging to its respective optimum numerically.

In total, we developed 18 distinct ISA solvers, all implemented in the \texttt{Horton-Part} package.\cite{YingXing2023}
These solvers include 1 GISA, 1 MBIS, 2 \clspg{LISA}, 2 \clspa{LISA}, 5 \clsng{LISA}, and 5 \clsna{LISA} solvers.

Regarding numerical benchmarks, we computed AIM charges for 42 organic and inorganic molecules in the TS42 database, along with six additional charged molecules from the literature, using these solvers.
Our initial results showed that two \clspa{LISA} (\m{aLISA-CVXOPT} and \m{aLISA-SC}), two \clsna{LISA} (\m{aLISA-M-NEWTON} and \m{aLISA-QUASI-NEWTON}), two \clspg{LISA} (\m{gLISA-CVXOPT} and \m{gLISA-SC}), and two \clsng{LISA} (\m{gLISA-FD-DIIS} and \m{gLISA-M-NEWTON}) converged for all molecules.
Our results demonstrate that the entropy obtained from \clsn{LISA} methods is, unsurprisingly, lower than that obtained from all \clsp{LISA} solvers.
However, \m{gLISA-FD-DIIS}, which lacks the restriction on the sign of the parameters, does not always obtain a solution with negative parameters corresponding to the lowest entropy.

We then employed the converged ISA solvers to compare their performance in terms of the number of outer iterations and total partitioning time.
Additionally, we investigated the entropy convergence behavior for each solver.
The key findings from our analysis are as follows:
\begin{enumerate}
    \item For \clspg{LISA} solvers, the \m{gLISA-SC} solver exhibited the highest number of outer iterations, making it the most computationally intensive in practice.
    Due to its lower number of outer iterations, the \m{gLISA-CVXOPT} solver was observed to be more efficient for small molecules than some \clsa{LISA} solvers.
    However, for larger molecules, it remained computationally demanding due to the need for gradient and Hessian matrices during optimization.
    \item Among \clsng{LISA} solvers, the \m{gLISA-M-NEWTON} solver demonstrated the best performance in terms of the number of outer iterations.
    Nevertheless, it was computationally expensive for larger molecules due to the cost of calculating the Jacobian matrix.
    The quasi-Newton method used by \m{gLISA-QUASI-NEWTON}, which generally provides lower entropy at a reduced computational cost, presents a potentially efficient and accurate AIM scheme.
    However, the numerical results indicate that it may not be robust for charged molecular systems, though it performed consistently well on neutral molecules in our test set.
    \item All \clsa{LISA} variants exhibited minor deviations in the outer iteration count across all systems, resulting in similar total times for all \clsa{LISA} methods, except for \m{aLISA-CVXOPT}, where the Hessian matrix is computationally expensive.
    \item After adding more diffuse basis functions for \ce{CH3-} and \ce{OH-}, all LISA-family solvers converged to the lowest entropy levels, while the \m{MBIS-SC} solver consistently converged to the highest entropy levels.
    All MBIS, \clspg{LISA}, and \clsa{LISA} solvers showed monotonic entropy decay.
    However, the \m{GISA-QUADPROG} solver exhibited non-monotonic entropy convergence with respect to outer iterations, consistent with previous studies.
    The \m{MBIS-SC} solver emerged as the fastest among the tested solvers, except for some \clsg{LISA} solvers (e.g., \m{gLISA-QUASI-NEWTON}), though it suffered from higher entropy values.
\end{enumerate}

Next, we compared the AIM charges obtained from the MBIS, GISA, and LISA-family methods.
We found that the difference in AIM charges obtained from the various LISA-family solvers was minimal, except for certain charged molecules.
Thus, we selected \m{aLISA-M-NEWTON} and \m{aLISA-SC} as representatives of all \clsna{LISA} and \clspa{LISA} solvers.
The results suggest that all \clsna{LISA} and \clspa{LISA} (i.e., \clsa{LISA}) solvers yield results that are more closely aligned with the ESP charges obtained by fitting both atomic charges and quadrupole moments, compared to those from the MBIS method.
Additionally, LISA solvers demonstrated superior performance in charged molecular systems compared to the MBIS method, particularly for AIM charges of anionic molecules.

In conclusion, this study provides a comprehensive evaluation of the performance of various ISA methods utilizing different solvers.
We have implemented computationally robust and efficient LISA-family variants, and our numerical results show that some even yield lower information entropy than MBIS, with reduced computational costs.

Table~\ref{tbl:summary} offers a comprehensive review and evaluation of various ISA methods, following the criteria outlined in Ref.~\citenum{Heidar-Zadeh2018}, and recommends solvers for each LISA category.
Considering the dependencies of the integration grids, \clsa{LISA} methods are preferable to \clsg{LISA} methods.
Therefore, the recommended solvers are \m{aLISA-SC} and \m{aLISA-M-NEWTON} for \clsp{LISA} and \clsn{LISA}, respectively.


\begin{table}[h]
\centering
\begin{threeparttable}
\caption{Checklist of desired AIM traits for different ISA methods.}
\begin{tabular*}{\textwidth}{@{\extracolsep{\fill}}lcccccc}
    \toprule
                                                 &               &               & \multicolumn{4}{c}{\textbf{LISA}} \\
                                                 \cmidrule{4-7}
                                                 &               &               & \multicolumn{2}{c}{\clsp{LISA}}& \multicolumn{2}{c}{\clsn{LISA}} \\
                                                 \cmidrule{4-5}
                                                 \cmidrule{6-7}
    \textbf{Features}                            & \textbf{GISA} & \textbf{MBIS} & \clspg{LISA}&\clspa{LISA}  &\clsng{LISA}&\clsna{LISA} \\
    \midrule
    \multicolumn{7}{c}{\textbf{Mathematical}}                                                                                              \\
    Universality                                 & \checkmark    & \checkmark    & \checkmark  &\checkmark    &\checkmark  &\checkmark     \\
    Information-Theoretic                        & \checkmark    & \checkmark    & \checkmark  &\checkmark    &\checkmark  &\checkmark     \\
    Variational $\rho(r)$                        & \texttimes    & \checkmark    & \checkmark  &\checkmark    &\checkmark  &\checkmark     \\
    Variational $\rho^0_a(r)$                    & \texttimes    & \checkmark    & \checkmark  &\checkmark    &\checkmark  &\checkmark     \\
    Uniqueness                                   & \texttimes    & \texttimes    & \checkmark  &\checkmark    &\checkmark  &\checkmark     \\
    \multicolumn{7}{c}{\textbf{Practical}}                                                                                                 \\
    Computational Robustness                     & \texttimes    & ?             & \checkmark  &\checkmark    &?           &\checkmark     \\
    Computational Efficiency                     & \checkmark    & \checkmark    & \checkmark  &\checkmark    &\checkmark  &\checkmark     \\
    \bottomrule
\end{tabular*}
\begin{tablenotes}
    \item[a] \checkmark~indicates the method complies with the feature.
    \item[b] \texttimes~indicates the method does not comply with the feature.
    \item[c] ?~indicates that further investigation is required.
    \item[d] Recommended \clspg{LISA} solver: \m{gLISA-CVXOPT}.
    \item[e] Recommended \clsng{LISA} solver: \m{gLISA-M-NEWTON}.
    \item[f] Recommended \clspa{LISA} solver: \m{aLISA-SC}.
    \item[g] Recommended \clsna{LISA} solver: \m{aLISA-M-NEWTON}.
\end{tablenotes}
\label{tbl:summary}
\end{threeparttable}
\end{table}

It is also important to note that the chemical accuracy and robustness of LISA-family methods have not been explored in this work.
These aspects are vital for practical applications and will be the focus of future studies.
This study, therefore, establishes a numerical groundwork for future research on LISA partitioning schemes in molecular density partitioning.

    \section*{Acknowledgements}
    \label{sec:acknowledgements}
    This project has received funding from the European Research Council (ERC) under the European Union's Horizon 2020 research and innovation program (grant agreement EMC2 No. 810367).
The resources and services used in this work were provided by the VSC (Flemish Supercomputer Center), funded by the Research Foundation - Flanders (FWO) and the Flemish Government.
We thank the Deutsche Forschungsgemeinschaft (DFG, German Research Foundation) for supporting this work by funding - EXC2075 – 390740016 under Germany's Excellence Strategy. We acknowledge the support by the Stuttgart Center for Simulation Science (SimTech).
AJM acknowledges funding from UKRI (Grant EP/X036863/1) for part of the PHYMOL (Physics, Accuracy and Machine-Learning: Towards the next-generation of Molecular-Potentials) MSCA Doctoral Network (Project 101073474).

    \section*{Dataset Availability Statement}
    \label{sec:data}
    The data that support the findings of this study are available within the article and its supplementary material.
The ``Horton-Part'' package can be found at \url{https://github.com/LISA-partitioning-method/horton-part}.

    \section*{Supplementary Material}
    \label{sec:suppmat}
    The Supplementary Material is a PDF document that includes the exponents and initial values for the Gaussian density basis used in both GISA and LISA methods, and the evolution of entropy for other molecules, which are not covered in the main text.

    \bibliography{references}
    \bibliographystyle{unsrt}

\end{document}